\documentclass[]{rsproca_new}
\jname{rspa}

\usepackage{tikz,xcolor,hyperref}

\definecolor{lime}{HTML}{A6CE39}
\DeclareRobustCommand{\orcidicon}{%
	\begin{tikzpicture}
	\draw[lime, fill=lime] (0,0) 
	circle [radius=0.16] 
	node[white] {{\fontfamily{qag}\selectfont \tiny ID}};
	\draw[white, fill=white] (-0.0625,0.095) 
	circle [radius=0.007];
	\end{tikzpicture}
	\hspace{-2mm}
}

\foreach \x in {A, ..., Z}{%
	\expandafter\xdef\csname orcid\x\endcsname{\noexpand\href{https://orcid.org/\csname orcidauthor\x\endcsname}{\noexpand\orcidicon}}
}

\begin{document}

\title{Kinematic dynamos in triaxial ellipsoids}

\author{J\'er\'emie Vidal and David C\'ebron}

\address{Universit\'e Grenoble Alpes, CNRS, ISTerre, 38000 Grenoble, France\\
\orcidA{ JV, \href{https://orcid.org/0000-0002-3654-6633}{0000-0002-3654-6633}}}

\subject{geophysics, fluid mechanics, applied mathematics}


\corres{J\'er\'emie Vidal\\\email{jeremie.vidal@univ-grenoble-alpes.fr}}
\esm{}

\begin{abstract}
Planetary magnetic fields are generated by motions of electrically conducting fluids in their interiors. 
The dynamo problem has thus received much attention in spherical geometries, even though planetary bodies are non-spherical. 
To go beyond the spherical assumption, we develop an algorithm that exploits a fully spectral description of the magnetic field in triaxial ellipsoids to solve the induction equation with local boundary conditions (i.e. pseudo-vacuum or perfectly conducting boundaries). 
We use the method to compute the free-decay magnetic modes and to solve the kinematic dynamo problem for prescribed flows. 
The new method is thoroughly compared with analytical solutions and standard finite-element computations, which are also used to model an insulating exterior. 
We obtain dynamo magnetic fields at low magnetic Reynolds numbers in ellipsoids, which could be used as simple benchmarks for future dynamo studies in such geometries.
We finally discuss how the magnetic boundary conditions can modify the dynamo onset, showing that a perfectly conducting boundary can strongly weaken dynamo action, whereas pseudo-vacuum and insulating boundaries often give similar results.
\end{abstract}


\begin{fmtext}
\section{Introduction}
Planetary magnetic fields are known to be generated by dynamo action, through complex motions of electrically conducting fluids in their liquid interiors. 
Many works have been devoted to convection-driven dynamos in spherical geometries, since these motions are commonly attributed to turbulent convection, and direct numerical simulations (DNS) have recently managed to reproduce some features of the geodynamo \cite{schaeffer2017turbulent,aubert2019approaching}.
However, the standard convective dynamo model cannot currently explain the origin of the magnetic fields of all planetary bodies (e.g. the Moon \cite{oran2020moon}). 
Thus, alternative models also have to be considered for planetary applications.
\end{fmtext}


\maketitle

Dynamo studies are usually performed in plane-layer, cylindrical or spherical geometries, where the mathematical problem can be efficiently solved using spectral methods. 
However, planetary bodies are non-spherical, for instance because of centrifugal effects or mechanical forcings (e.g. tides). 
The simplest non-spherical geometry that is relevant for planetary bodies is the ellipsoid, and so flows driven by mechanical forcings in ellipsoids have received a renewed interest in the fluid community \cite{le2015flows}. 
The forced laminar response of a fluid-filled ellipsoid to a mechanical forcing is a priori not dynamo capable for planetary-like parameters (e.g. \cite{loper1975torque,rochester1975can,tilgner1998models} for precession). 
However, mechanical forcings can sustain flow instabilities in rotating ellipsoids \cite{kerswell2002elliptical,vantieghem2015latitudinal,vidal2017inviscid}, which could lead to space-filling turbulence \cite{horimoto2018impact,le2019experimental} and possibly large-scale dynamo fields (e.g. as reported for ad hoc tidal forcings \cite{cebron2014tidally,vidal2018magnetic}). 
Therefore, ellipsoidal geometries should be considered to explore mechanically driven dynamos for planetary applications.

Given the recent availability of massive computations, it is sometimes argued that only (nonlinear) saturated dynamos are worth considering to make progress in dynamo theory \cite{tobias2021turbulent}. 
This could be true in plane-layer or spherical domains, but solving the dynamo problem in triaxial ellipsoids is strongly hampered by the mathematical complexity of the geometry. 
Spectral decompositions of vector fields in non-orthogonal coordinates have been proposed \cite{schmitt2004numerical,ivers2017kinematic}, but the numerical implementation is very challenging. 
Non-spectral numerical methods have also been employed, for instance using local methods that can accommodate more easily non-spherical geometries (e.g. finite volumes \cite{ernst2013finite,vantieghem2016applications}, finite elements \cite{cebron2012magnetohydrodynamic}, or spectral elements \cite{guermond2009nonlinear,reddy2018turbulent}).
Yet, numerical convergence is slower to achieve with local methods than with spectral methods (see the international comparisons \cite{jackson2014spherical,matsui2016performance} in spherical geometries). 
The reliability of the previously published dynamo solutions in ellipsoids is also questionable. 
For instance, the dynamo solutions driven by topographic precession or librations in spheroids \cite{wu2009dynamo,wu2013dynamo} cannot be replicated numerically (because the velocity boundary conditions led to spurious behaviours \cite{guermond2013remarks}). 
Similarly, the kinematic dynamos in ellipsoids presented in \cite{reddy2018turbulent} could be very difficult to replicate because the considered dynamo-capable flows are not laminar (contrary to the convection-driven dynamo benchmarks). 
Therefore, given the computational burden in ellipsoids, it is essential to provide dynamo solutions that could be easily reproduced by any dynamo codes. 

To do so, the first step is to consider the kinematic dynamo problem for computational simplicity, where the velocity field is prescribed. 
A huge amount of effort has been devoted to this problem in spherical geometries, showing that it is difficult to obtain numerical solutions of the kinematic induction equation. 
Some early models have indeed proven unconvincing \cite{bullard1954homogeneous,kumar1975three}, owing to insufficient numerical resolution \cite{pekeris1973kinematic} or mathematical inconsistencies in the description of the magnetic field \cite{dudley1989time}. 
The ability to reproduce some kinematic solutions in DNS has also been called into question, because very small differences in the flow could drastically change the linear stability results \cite{livermore2004magnetic}. 
It is thus clear that extreme care must be taken to ensure numerical convergence of dynamo computations, because insufficient resolution would otherwise favour spurious dynamo fields. 
A careful investigation of the kinematic dynamo problem remains to be undertaken in ellipsoidal geometries, to obtain robust dynamo solutions for future numerical validations in triaxial ellipsoids (prior to simulations of saturated dynamos). 

In the present study, we aim to propose robust simple solutions of the kinematic dynamo problem in full ellipsoids.  
Such geometries are also directly relevant for some planetary bodies (e.g. the primitive Earth \cite{monville2019rotating} or the Moon \cite{viswanathan2019observational}). 
Fully spectral algorithms that rely on global polynomial descriptions satisfying the boundary conditions have already proven accurate for this problem in spheres \cite{livermore2005comparison,li2010optimal,livermore2010galerkin}, but such methods remained to be devised for dynamos in ellipsoids. 
We thus develop here a fully spectral algorithm in triaxial ellipsoids, motivated by our previous hydrodynamic works in ellipsoids \cite{vidal2020compressible,vidal2020acoustic}. 
To sidestep the known difficulties of the ellipsoidal coordinate system, we employ the Cartesian coordinates and expand the magnetic field onto global Cartesian polynomial elements satisfying local boundary conditions (BC), namely pseudo-vacuum or perfectly conducting BC.
Using such BC in non-spherical geometries is indeed simpler than using insulating BC, which are not straightforward to implement in non-spherical dynamo codes \cite{iskakov2004integro,ivers2017kinematic}.   
The paper is divided as follows. 
We describe the dynamo problem in \S\ref{sec:problem}, and the numerical methods in \S\ref{sec:numerics}. 
Numerical results are presented in \S\ref{sec:results}, and some implications for planetary modelling are discussed in  \S\ref{sec:discussion}. 
We conclude the paper in \S\ref{sec:ccl}.

\section{Formulation of problem}
\label{sec:problem}
\subsection{Kinematic dynamo problem}
We consider an impermeable triaxial ellipsoid of semi-axes $[a,b,c]$ and volume $V$, surrounded by a quiescent exterior region. 
The ellipsoid is filled with a Newtonian and electrically conducting fluid of uniform magnetic diffusivity $\eta$. 
We employ throughout the paper the Cartesian coordinates $(x,y,z)$, and denote $\boldsymbol{r} = (x,y,z)^\top$ as the position vector.
We work in a reference frame where the ellipsoidal boundary $\partial V$ is steady, and given by the quadratic equation $F=1$ with $F = (x/a)^2 + (y/b)^2 + (z/c)^2$.
We make the kinematic dynamo approximation below, which neglects the feedback of the magnetic field onto the flow.
The fluid is assumed to move at the prescribed (here steady) velocity field $\boldsymbol{v}$, which is divergenceless $\nabla \boldsymbol{\cdot} \boldsymbol{v} = 0$ (as for incompressible fluids) and obeys the no-penetration boundary condition (BC) $\left . \boldsymbol{v} \boldsymbol{\cdot} \boldsymbol{n} \right|_{\partial V} = 0$ on the boundary $\partial V$ (where $\boldsymbol{n} = (x/a^2, y/b^2, z/c^2)^\top$ is the non-unit outward normal vector at the boundary). 

The time evolution of the magnetic field $\boldsymbol{B}$ is governed by the induction equation \cite{backus1996foundations}
\begin{subequations}
\label{eq:inductionB}
\begin{equation}
    \partial_t \boldsymbol{B} = \nabla \times (\boldsymbol{v} \times \boldsymbol{B}) + \eta \nabla^2 \boldsymbol{B}, \quad \nabla \boldsymbol{\cdot} \boldsymbol{B} = 0.
    \tag{\theequation \emph{a,b}}
\end{equation}
\end{subequations}
We also introduce the magnetic potential vector $\boldsymbol{A}$, defined by $\boldsymbol{B} = \nabla \times \boldsymbol{A}$. 
The induction equation can then be written in the alternative form \cite{cebron2012magnetohydrodynamic} 
\begin{equation}
    \partial_t \boldsymbol{A} = \boldsymbol{v} \times (\nabla \times \boldsymbol{A}) +  \eta \, \nabla^2 \boldsymbol{A},
    \label{eq:inductionA}
\end{equation}
where we have employed the Weyl gauge (such that $\boldsymbol{A}$ is not divergenceless). 
Formulation (\ref{eq:inductionA}) ensures that the magnetic field is solenoidal with the standard finite-element method, which will be used below for numerical validation (see appendix \ref{appendix:fem}). 

\subsection{Magnetic boundary conditions}
Two sets of local BC are examined here.  
We consider the pseudo-vacuum BC (PV BC) \cite{vantieghem2016applications}
\begin{subequations}
\label{eq:BCferro}
\begin{equation}
    \left . \boldsymbol{B} \times \boldsymbol{n} \right|_{\partial V} = \boldsymbol{0} \quad \Longrightarrow \quad \left . (\nabla \times \boldsymbol{B}) \boldsymbol{\cdot} \boldsymbol{n} \right|_{\partial V} = 0,
    \tag{\theequation \emph{a,b}}
\end{equation}
\end{subequations}
which notably require that the magnetic field is purely normal at the boundary. 
Using PV BC (\ref{eq:BCferro}) is a reasonable assumption when the magnetic permeability of the exterior is much larger than the fluid magnetic permeability $\mu$ (e.g. in laboratory experiments \cite{gissinger2008effect}), and PV BC are also often used for stellar modelling \cite{jouve2008solar}. 

Alternatively, we can assume that the exterior is a perfect electrical conductor to use the perfectly conducting BC (PC BC) \cite{cebron2012magnetohydrodynamic}
\begin{subequations}
\label{eq:BCsupra1}
\begin{equation}
    \left . \boldsymbol{A} \times \boldsymbol{n} \right|_{\partial V} = \boldsymbol{0} \quad  \Longrightarrow \quad
    \left . \boldsymbol{B} \boldsymbol{\cdot} \boldsymbol{n} \right|_{\partial V} = 0.
    \tag{\theequation \emph{a,b}}
\end{equation}
\end{subequations}
PC BC (\ref{eq:BCsupra1}) are often used in diffusionless models, for instance to study Alfv\'en waves \cite{malkus1967hydromagnetic,kerswell1994tidal,labbe2015magnetostrophic} or hydromagnetic instabilities \cite{kerswell1994tidal,zhang2003nonaxisymmetric}.
In the presence of magnetic diffusion, Ohm's law also provides the additional BC 
\begin{equation}
    \left . (\nabla \times \boldsymbol{B}) \times \boldsymbol{n} \right|_{\partial V} = \boldsymbol{0}.
    \label{eq:BCsupra2}
\end{equation}
Finally, the observed similarity between BC (\ref{eq:BCferro}) and BC (\ref{eq:BCsupra1}) indicates that there is some kind of duality between these two BC, as formally demonstrated in the presence of velocity fields that obey some symmetries  \cite{favier2013growth}. 

\subsection{Dimensionless form}
The dynamo equation is here non-dimensionalised by using the typical equatorial radius $R=\sqrt{(a^2+b^2)/2}$ as the length scale, the magnetic diffusion time $R^2/\eta$ as the time scale, and $\eta/R$ as the velocity scale. 
Non-dimensionalizing the ellipsoidal geometry yields the dimensionless equatorial semi-axes $\tilde{a} = a/R=\sqrt{1+\beta}$ and $\tilde{b} = b/R=\sqrt{1-\beta}$ with the equatorial ellipticity $\beta = |a^2-b^2|/(a^2+b^2)$, and the polar semi-axis $\tilde{c} = c/R$.
We employ dimensionless variables in the following and, for the sake of conciseness, we do not introduce specific symbols to distinguish the other dimensionless variables from their dimensional counterparts. 

The time evolution of the magnetic field is basically governed by a competition between advection, which can act to intensify the magnetic field, and Ohmic diffusion that destroys it.
Accordingly, the magnetic field will decay in time unless the amplitude of the induction term $|\nabla \times (\boldsymbol{v} \times \boldsymbol{B})|$ is large enough compared with that of Ohmic diffusion $|\eta \nabla^2 \boldsymbol{B}|$. 
To compare these two effects, we generally introduce the energy-based magnetic Reynolds number
\begin{subequations}
\label{eq:RmU}
\begin{equation}
    Rm = \frac{U \, R}{\eta}, \quad U = \sqrt{\frac{1}{V} \int |\boldsymbol{v}|^2 \, \mathrm{d}V},
    \tag{\theequation \emph{a,b}}
\end{equation}
\end{subequations}
where $U$ is a volume-averaged amplitude for the velocity. 
Large values $Rm~\gg~1$ mean that advection dominates over diffusion, and solving the kinematic dynamo problem often amounts to finding the minimum value of $Rm$ for dynamo action. 
However, the previous definition of $Rm$ is not always appropriate to estimate the relative importance of advection and diffusion. 
It is indeed sometimes possible to reduce the size of the fluid domain and find some dynamo capable flows when $Rm \to 0$ \cite{proctor2015energy}, but the standard definition is still commonly used.

\section{Spectral algorithm}
\label{sec:numerics}
Numerical solutions of partial differential equations can often be obtained using Galerkin projection methods \cite{finlayson1972method}. 
Such methods require the unknown field to be expanded onto a linear combination of basis elements satisfying the BC. 
Here, it is advantageous to construct the basis elements for the magnetic field using spectral decompositions that admit explicit polymonial expressions (e.g. as in spheres \cite{livermore2005comparison,li2010optimal,luo2020optimal}). 
We first introduce suitable spectral decompositions in triaxial ellipsoids, and then describe the projection method.

\subsection{Polynomial expansions}
To account for PV BC, we write the magnetic field $\boldsymbol{B}$ and $\nabla \times \boldsymbol{B}$ in the form
\begin{subequations}
\label{eq:spectralab}
\begin{equation}
    \boldsymbol{B} = \nabla \times (\mathcal{A} \, \boldsymbol{n}) +  \mathcal{B} \, \boldsymbol{n} + \nabla \Phi, \quad \nabla \times \boldsymbol{B} = \nabla \times \nabla \times (\mathcal{A} \, \boldsymbol{n}) + \nabla \times (\mathcal{B} \, \boldsymbol{n}),
    \tag{\theequation \emph{a,b}}
\end{equation}
\end{subequations}
where $[\mathcal{A},\mathcal{B}]$ are poloidal-toroidal scalars and $\Phi$ is a scalar potential.
We also impose
\begin{equation}
    \left [ \boldsymbol{n} \boldsymbol{\cdot} \nabla + \nabla \boldsymbol{\cdot} \boldsymbol{n} \right ] \, \mathcal{B} = -\nabla^2 \Phi
    \label{eq:diva}
\end{equation}
to enforce the divergenceless condition (\ref{eq:inductionB}b), and PV BC are automatically satisfied if $\left . \mathcal{A} \right|_{\partial V} = \left .  \Phi\right|_{\partial V} = 0$ on the boundary. 
For PC BC, we instead write the magnetic field in the form 
\begin{equation}
     \boldsymbol{B} = \nabla \times \left ( \Lambda \nabla \Upsilon  \right ) = \nabla \Lambda \times \nabla \Upsilon
    \label{eq:clebsch}
\end{equation}
where $[\Lambda, \Upsilon]$ are Clebsch (or Euler) potentials \cite{marques1990boundary}.
As uncovered by Lebovitz \cite{lebovitz1989stability}, decomposition (\ref{eq:clebsch}) exactly enforces PC BC (\ref{eq:BCsupra1}) if either $\Lambda$ or $\Upsilon$ is constant everywhere on the boundary. 
This mathematical decomposition is exact without magnetic diffusion, but does not allow the magnetic field to exactly obey BC (\ref{eq:BCsupra2}).
The latter BC must indeed be fulfilled if the magnetic diffusion is non-zero in the ellipsoidal interior.
As within the standard finite-element method, this additional BC is thus enforced in the weak formulation of the diffusion term (see the next subsection).

We can then approximate the magnetic field by constructing polynomial bases made of vector elements in the form of Cartesian monomials of maximum degree $N$, and which satisfy either PV BC or PC BC.
To do so, we introduce the finite-dimensional space $\mathcal{P}_N$, which is made of scalar polynomials in $(x,y,z)$ and of degree $N$ or less. 
The magnetic field elements are of maximum polynomial degree $N$ if $\mathcal{A} \in \mathcal{P}_N$, $\mathcal{B} \in \mathcal{P}_{N-1}$ and $\Phi \in \mathcal{P}_{N+1}$ in decomposition (\ref{eq:spectralab}) for PV BC, and if $\Lambda + \Upsilon \in \mathcal{P}_{N+2}$ for PC BC.
The explicit forms of the scalars in ellipsoids are given in appendix \ref{appendix:basis} for PV BC, and in \cite{lebovitz1989stability,vidal2020acoustic} for PC BC.

\subsection{Projection method}
Because the velocity field is assumed to be steady, we seek solutions for the magnetic field $\boldsymbol{B}$ 
using the finite-dimensional expansion
\begin{subequations}
\label{eq:trial}
\begin{equation}
    \boldsymbol{B}(\boldsymbol{r},t) = \sum_{j = 1}^{\mathcal{N}} \gamma_j \, \boldsymbol{e}_j (\boldsymbol{r}) \exp (\lambda t), \quad \nabla \boldsymbol{\cdot} \boldsymbol{e}_j = 0,
    \tag{\theequation \emph{a,b}}
\end{equation}
\end{subequations}
where $\boldsymbol{\gamma} = (\gamma_1, \gamma_2, \dots, \gamma_\mathcal{N})^\top$ is the state vector, $\{\boldsymbol{e}_j \}_{1\leq j\leq \mathcal{N}}$ are the real-valued polynomial elements that satisfy the BC, and $\lambda = \sigma + \mathrm{i} \omega$ is the complex-valued eigenvalue with the growth rate $\sigma \geq 0$ (or decay rate when $\sigma < 0$) and the angular frequency $\omega\in \mathbb{R}$ of the magnetic field.

\begin{figure}
    \centering
    \begin{tabular}{ccc}
    \includegraphics[width=0.31\textwidth]{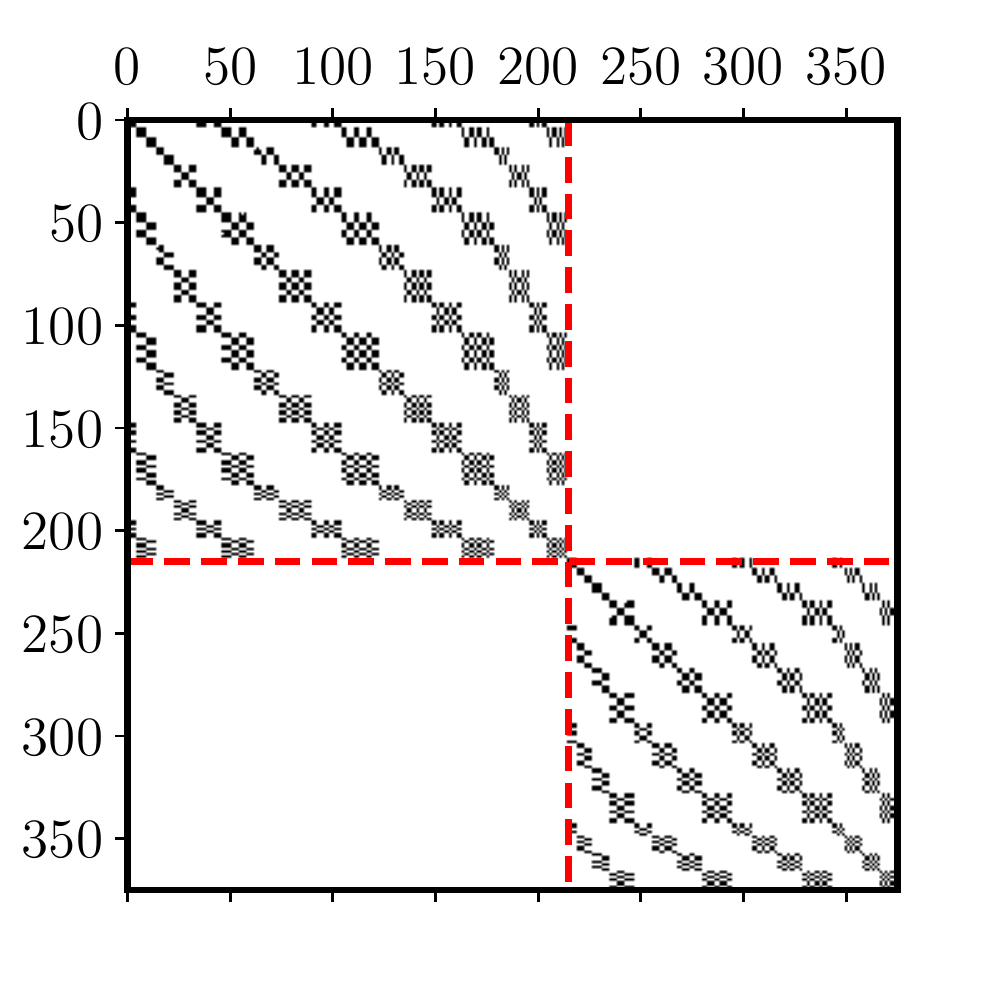} & 
    \includegraphics[width=0.31\textwidth]{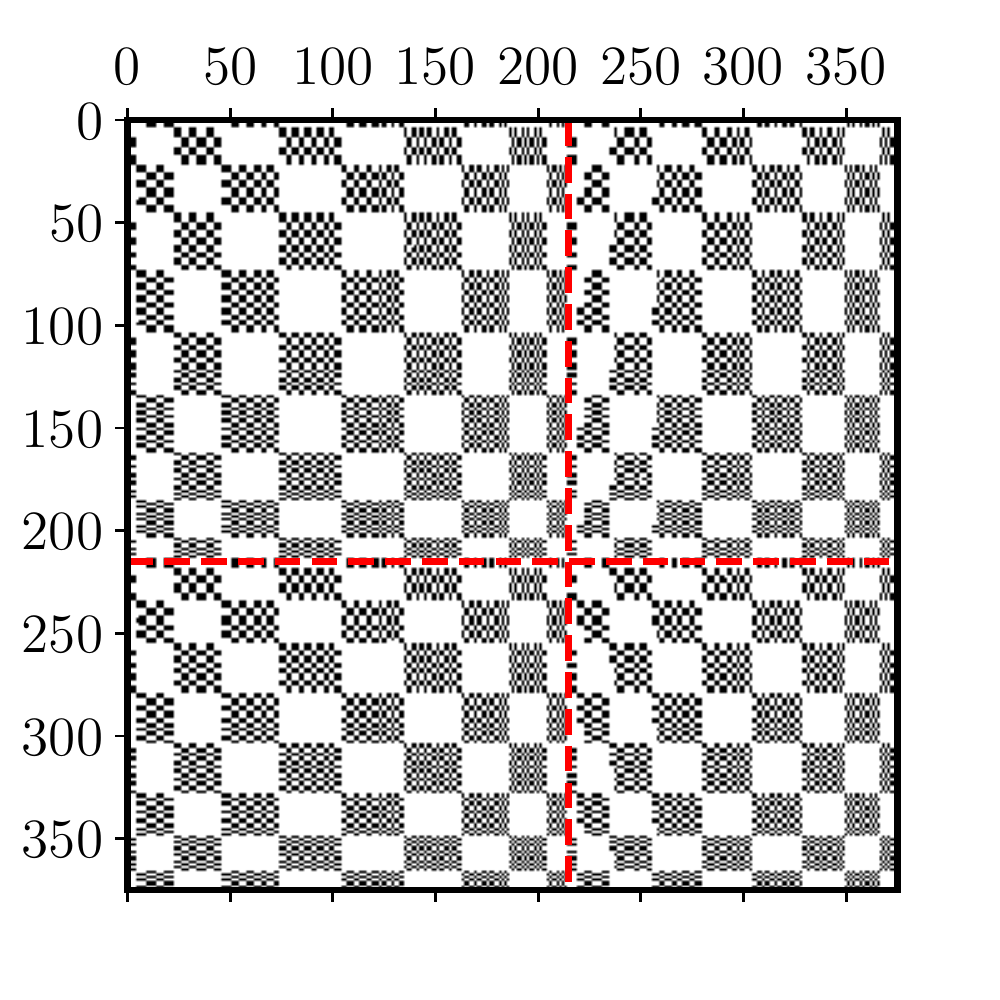} &
    \includegraphics[width=0.31\textwidth]{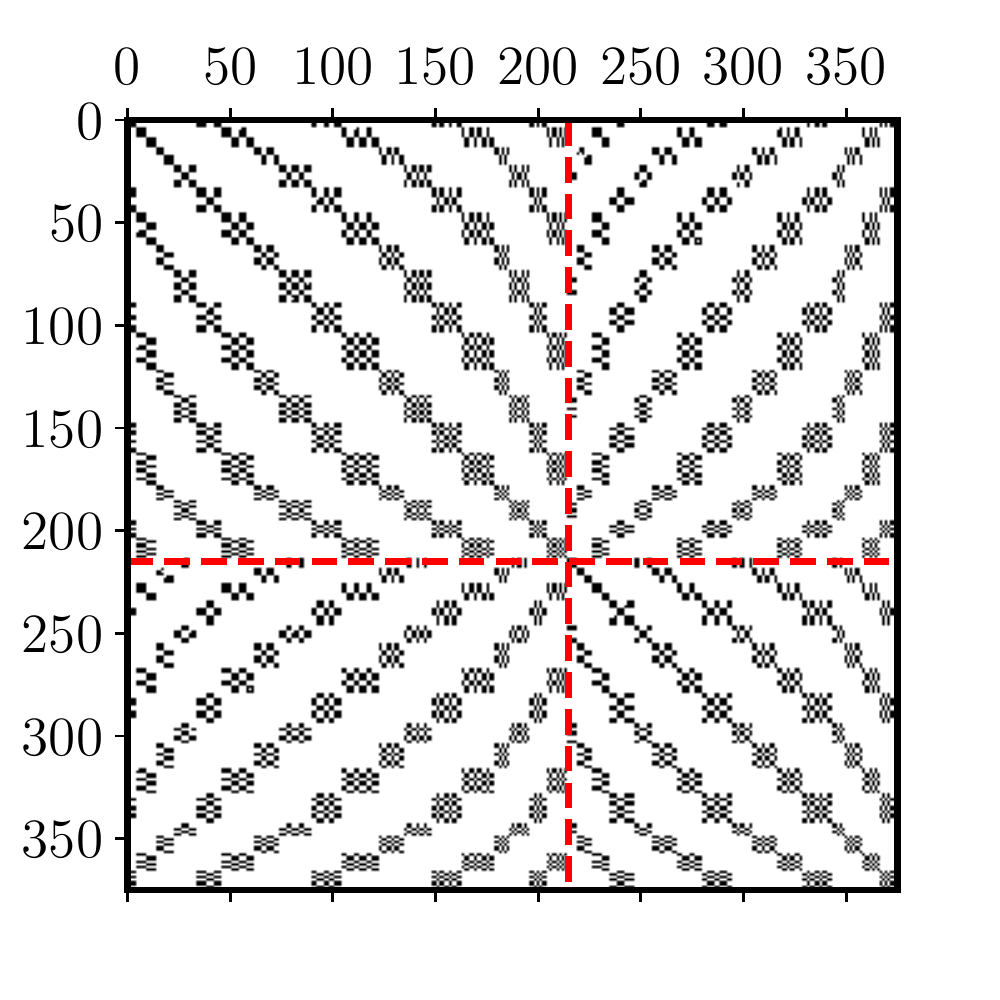} \\
    $\boldsymbol{L}$ & $\boldsymbol{D}$ & $\boldsymbol{R}$ \\
    \end{tabular}
    \caption{Non-zero elements (\ref{eq:Galerkinproj}) of matrices $[\boldsymbol{L},\boldsymbol{D},\boldsymbol{R}]$ with PV BC, computed at polynomial degree $N=10$ for flow $T_1^0 P_1^0$ with $\epsilon_1=\epsilon_2=1$. The number of basis elements is $\mathcal{N}=375$. Thick dashed lines indicate the block structure associated with the two vector components $\mathcal{B} \, \boldsymbol{n}+\nabla \Phi$ and $\nabla \times (\mathcal{A} \boldsymbol{n})$ in decomposition (\ref{eq:spectralab}). (Online version in colour.)}
    \label{fig:matrices}
\end{figure}

We then substitute expansion (\ref{eq:trial}) into the induction equation, and project it onto every basis element $\boldsymbol{e}_i$ to minimise the residual terms with respect to the inner product $\langle \boldsymbol{e}_i,\boldsymbol{e}_j \rangle = \int \boldsymbol{e}_i^\dagger \boldsymbol{\cdot} \boldsymbol{e}_j \ \mathrm{d}V$ (where ${}^\dagger$ denotes the complex conjugate). 
We obtain the dimensionless eigenvalue problem
\begin{equation}
    \lambda \, \boldsymbol{L} \, \boldsymbol{\gamma} = \left [\boldsymbol{R} - \boldsymbol{D} \right ] \, \boldsymbol{\gamma}, 
    \label{eq:MatrixDynamoPb}
\end{equation}
with the three real-valued matrices $[\boldsymbol{L}, \boldsymbol{R}, \boldsymbol{D}]$ of non-zero elements
\begin{subequations}
\label{eq:Galerkinproj}
\begin{equation}
     \boldsymbol{L}_{ij} = \langle \boldsymbol{e}_i, \boldsymbol{e}_j \rangle, \quad \boldsymbol{R}_{ij} = \langle \boldsymbol{e}_i, \nabla \times (\boldsymbol{v} \times \boldsymbol{e}_j) \rangle, \quad \boldsymbol{D}_{ij} = - \langle \boldsymbol{e}_i, \nabla^2 \boldsymbol{e}_j \rangle = \langle \nabla \times \boldsymbol{e}_i,\nabla \times \boldsymbol{e}_j \rangle.
    \tag{\theequation \emph{a--c}}
\end{equation}
\end{subequations}
Note that we have used PV BC (\ref{eq:BCferro}) or PC BC (\ref{eq:BCsupra1})-(\ref{eq:BCsupra2}) to rewrite the volume integral in expression (\ref{eq:Galerkinproj}c) in a symmetric form. 
This symmetric formulation thus ensures that the local BC are satisfied in the weak form of the problem. 
The matrix structure is illustrated in figure \ref{fig:matrices}.
The matrix $\boldsymbol{L}$ is sparse and symmetric positive definite (SPD),  $\boldsymbol{D}$ is SPD (but only with PV BC or PC BC), and $\boldsymbol{R}$ is known to be strongly non-normal. 
We only consider in the following velocity fields $\boldsymbol{v}$ that have exact polynomial components in the Cartesian coordinates, such that projections (\ref{eq:Galerkinproj}) can be evaluated analytically (see formula (50) in \cite{lebovitz1989stability}). 
We have modified our bespoke code \cite{vidal2020acoustic} to implement the above algorithm for the dynamo problem.

\section{Results}
\label{sec:results}
We present below some numerical results, which could be used as a benchmark for future dynamo studies in ellipsoids. 
We first assess the convergence of the polynomial description by computing the free-decay magnetic modes, and then solve the kinematic dynamo problem for some prescribed flows. 
To ease future validation of dynamo codes in ellipsoidal geometries, we have gathered in table \ref{table:bench} some benchmark values for the different cases studied in this article.

\begin{table}
    \centering
    \caption{Selected benchmark values for the growth rate $\sigma \geq 0$ (or decay rate $\sigma < 0$) and angular frequency $|\omega|$ of the fastest growing (or slowest decaying) magnetic field solution of the kinematic dynamo equation for flows $[T_1^0 P_1^0,T_1^0 P_2^0]$ (or free-decay modes, symbol $\emptyset$), for an ellipsoid of semi-axes $[\sqrt{1+\beta},\sqrt{1-\beta},\tilde{c}]$ with pseudo-vacuum (PV), perfectly conducting (PC) or insulating (IN) boundary conditions (BC). 
    Dimensionless amplitude of the toroidal and poloidal components are respectively given by $\epsilon_1$ and $\epsilon_2$. The calculations using the finite-element method (FEM), performed with \textsc{comsol}, typically require $5 \times 10^5- 10^6$ degrees of freedom to ensure good numerical convergence.}
    \begin{tabular}{cccccccccc}
    \hline
    Flow & $\tilde{c}$ & $\beta$ & BC & $\epsilon_1$ & $\epsilon_2$ & $\sigma$ & $|\omega|$ & Method \\
    \hline
    $T_1^0 P_1^0$ & $0.95$ & $0$ & PV & $210$ & $120$ & $-5.154$ & $0$ & FEM \\
    $T_1^0 P_1^0$ & $0.95$ & $0.1$ & PV & $210$ & $120$ & $-6.271$ & $0$ & FEM  \\
    $T_1^0 P_1^0$ & $0.95$ & $0.5$ & PV & $210$ & $120$ & $4.801$ & $39.34$ & FEM  \\
    $T_1^0 P_1^0$ & $0.95$ & $0$ & IN & $210$ & $120$ & $-6.9596$ & $0$ & FEM  \\
    $T_1^0 P_1^0$ & $0.95$ & $0.1$ & IN & $210$ & $120$ & $-8.633$ & $0$ & FEM \\
    $T_1^0 P_1^0$ & $0.95$ & $0.5$ & IN & $210$ & $120$ & $1.822$ & $40.67$ & FEM \\
    $T_1^0 P_1^0$ & $0.95$ & $0.6$ & IN & $210$ & $120$ & $3.431$ & $30.75$ & FEM \\
    \hline
    $T_1^0 P_2^0$ & $1$ & $0.1$ & PV & $190$ & $35$ & $0.9776$ &  $31.90$ & FEM \\
    $T_1^0 P_2^0$ & $1$ & $0.2$ & PV & $190$ & $35$ & $0.948$ &  $32.30$ & FEM \\
    $T_1^0 P_2^0$ & $1$ & $0.6$ & PV & $190$ & $35$ & $0.307$ &  $37.51$ & FEM \\
    $T_1^0 P_2^0$ & $1$ & $0.1$ & IN & $190$ & $35$ & $-1.729$ &  $32.39$ & FEM \\
    $T_1^0 P_2^0$ & $1$ & $0.2$ & IN & $190$ & $35$ & $-1.824$ &  $32.78$ & FEM \\
    $T_1^0 P_2^0$ & $1$ & $0.6$ & IN & $190$ & $35$ & $-3.149$ &  $38.1$ & FEM  \\
    $T_1^0 P_2^0$ & $1$ & $0.44$ & PC & $860$ & $137$ & $1.751$ &  $141.8$  & FEM \\
    $T_1^0 P_2^0$ & $1$ & $0.44$ & PC & $790$ & $110$ & $0.4118$ &  $159.8$ & FEM \\
    \hline
    $\emptyset$ & $0.4$ & $0$ & PV-PC & $0$ & $0$ & $-7.998$ &  $0$ & Spectral \\
    $\emptyset$ & $0.8$ & $0$ & PV-PC & $0$ & $0$ & $-7.696$ &  $0$ & Spectral \\
    $\emptyset$ & $1.2$ & $0$ & PV-PC & $0$ & $0$ & $-6.429$ &  $0$ & Spectral \\
    $\emptyset$ & $0.4$ & $0.44$ & PV-PC & $0$ & $0$ & $-9.831$ &  $0$ & Spectral \\
    $\emptyset$ & $0.6$ & $0.44$ & PV-PC & $0$ & $0$ & $-8.655$ &  $0$ & Spectral \\
    $\emptyset$ & $1.2$ & $0.44$ & PV-PC & $0$ & $0$ & $-5.445$ &  $0$ & Spectral \\
    \hline
    \end{tabular}
    \label{table:bench}
\end{table}

\subsection{Free-decay magnetic modes}
\begin{figure}
    \centering
    \begin{tabular}{cc}
    \includegraphics[width=0.49\textwidth]{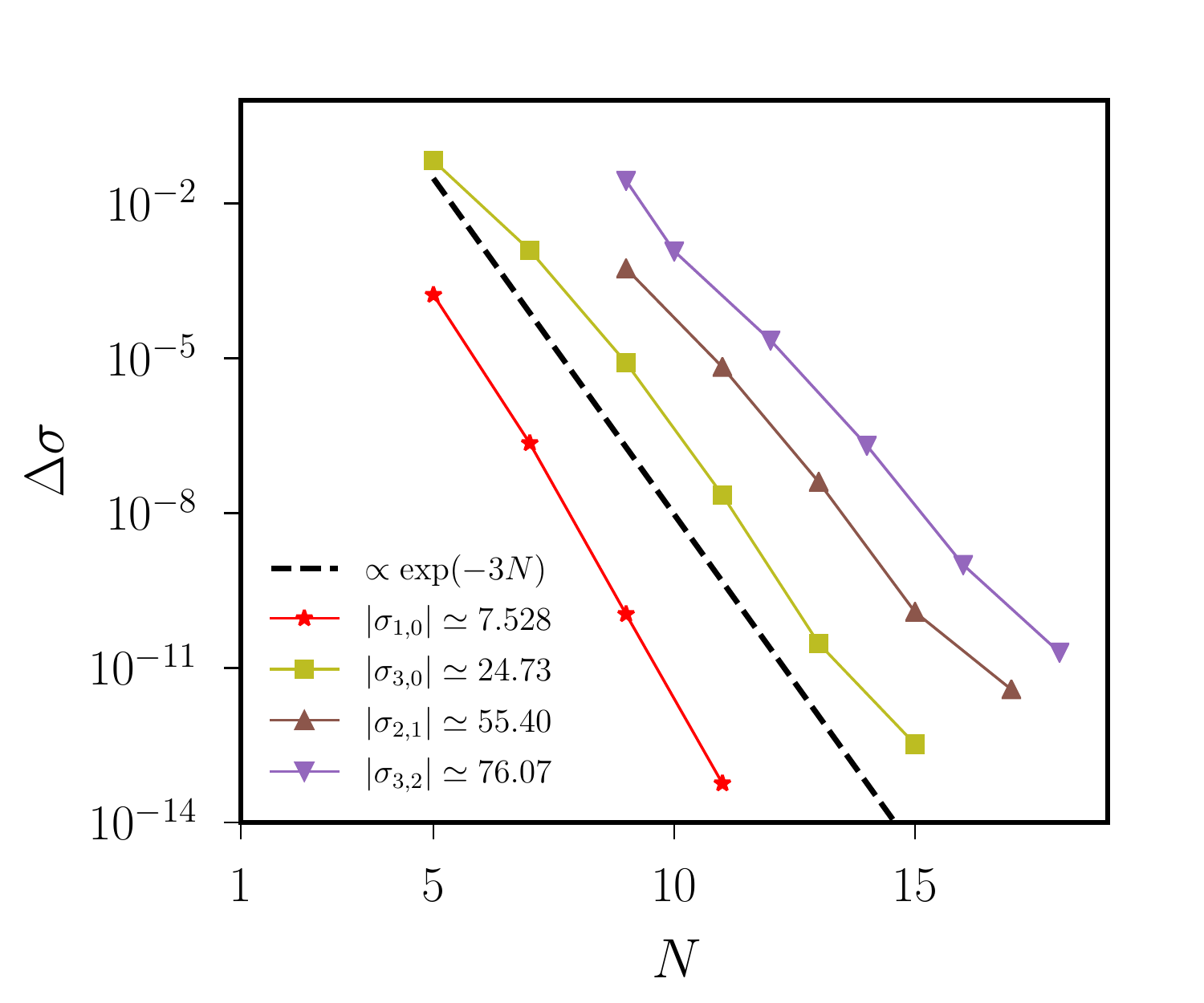} &
    \includegraphics[width=0.49\textwidth]{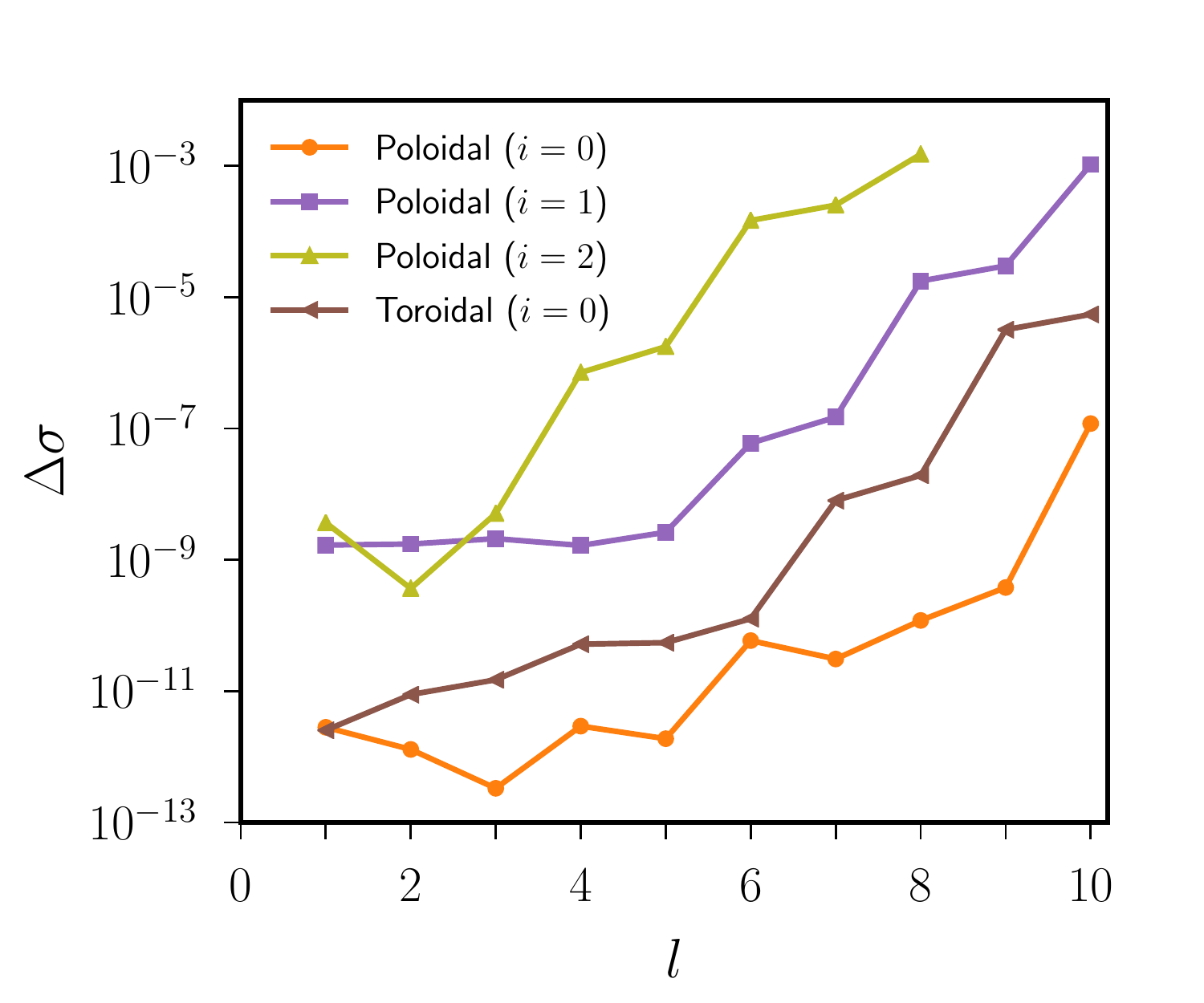} \\
    (a) & (b)
    \end{tabular}
    \caption{Numerical convergence of free-decay magnetic modes with PV BC in spheres. (a) Relative error $\Delta \sigma$ as a function of polynomial degree $N$ for a few poloidal modes. Dashed line indicates the spectral convergence $\Delta \sigma \propto \exp(-3 N)$.
    (b) $\Delta \sigma$ as a function of the spherical harmonic degree $l$ of a few modes computed at $N=20$. (Online version in colour.)}
    \label{fig:freedecay}
\end{figure}

The free-decay magnetic modes are the eigensolutions of the induction equation with $\boldsymbol{v} = \boldsymbol{0}$ and $\lambda = \sigma < 0$ (because the problem is self-adjoint). 
We first consider a full sphere with a unit radius (i.e. $\tilde{a}=\tilde{b}=\tilde{c}=1$), since analytical solutions can be obtained in this geometry. 
We expand $\boldsymbol{B}$ onto the spherical harmonics $\mathcal{Y}_l^m$ of degree $l\geq1$ and order $|m|\leq l$ in the form
\begin{equation}
\boldsymbol{B} = \sum_i \sum_{1\leq l} \sum_{|m|\leq l} \left [ \nabla \times \nabla \times (P_{l,i}^m  \, \mathcal{Y}_l^m \, \boldsymbol{r}) + \nabla \times (T_{l,i}^m \, \mathcal{Y}_l^m \, \boldsymbol{r}) \right ] \exp(\lambda t),
\end{equation}
where the index $i$ accounts for the radial complexity of the modes. 
The poloidal-toroidal scalars for every pair $(l,m)$ are then given in dimensionless form by
\begin{subequations}
\begin{equation}
    \sigma_{l,i} P_{l,i}^m = \nabla^2 P_{l,i}^m, \quad \sigma_{l,i} T_{l,i}^m = \nabla^2 T_{l,i}^m,
    \tag{\theequation \emph{a,b}}
\end{equation}
\end{subequations}
noting that all the modes with $|m|\leq l$ have the same eigenvalue $\lambda = \sigma_{l,i}$. 
PV BC (\ref{eq:BCferro}) reduce to $\partial_r (r P_{l,i}^m) = T_{l,i}^m = 0$ at the outer radius $r=1$. 
We obtain the analytical solutions $[P_{l,i}^m , T_{l,i}^m] \propto (1/\sqrt{r}) \, \mathrm{J}_{l+1/2} (k_{l,i} r)$ with $(k_{l,i})^2 = -\sigma_{l,i}$, where $\mathrm{J}_{l+1/2}$ is the Bessel function of the first kind.
The BC discretised the allowed values for the decay rates of the poloidal and toroidal scalars, which are obtained respectively from the $i$-th roots of
\begin{subequations}
\label{eq:rootsbessel}
\begin{equation}
  k_{l,i} \, \mathrm{J}_{l-1/2}(k_{l,i}) = l \mathrm{J}_{l+1/2} (k_{l,i}), \quad \text{or} \quad \mathrm{J}_{l+1/2}(k_{l,i}) = 0.
  \tag{\theequation \emph{a,b}}
\end{equation}
\end{subequations}
Equations (\ref{eq:rootsbessel}) can be accurately solved using a nonlinear solver. 
The lowest zero of (\ref{eq:rootsbessel}a) with $l=1$ gives the slowest decaying poloidal mode with $\sigma_{1,0} \simeq -7.52793$, and that of (\ref{eq:rootsbessel}b) with $l=1$ gives the slowest decaying toroidal mode with $\sigma_{1,0} \simeq -20.19073$. 
The same analysis can be conducted for PC BC, and the BC take the form $\partial_r (r T_{l,i}^m)= P_{l,i}^m = 0$ on $r=1$.
It is thus obvious that the decay rates are identical between PV BC and PC BC \cite{favier2013growth}, except that the poloidal modes are exchanged with the toroidal modes.  

To assess the resolution of the polynomial solutions, it is worth looking at the convergence of the numerical decay rate $\sigma$, which is tightly related to the convergence of the spatial structure of the modes \cite{valdettaro2007convergence}.
We show in figure \ref{fig:freedecay}a the evolution of relative error $\Delta \sigma = |\sigma - \sigma_{l,i}|/|\sigma_{l,i}|$, as a function of the maximum polynomial degree $N$, for a few free-decay modes.  
The polynomial description is characterised by a spectral convergence with errors decreasing as $ \Delta \sigma \propto \exp(-\alpha N)$, where $1 \leq \alpha \leq 5$ is found to be mode dependent.
Such an exponential convergence is much faster than the standard algebraic convergence obtained with finite differences or finite elements (e.g. see appendix A in \cite{vidal2020compressible} for a different problem).
We also find that numerical convergence hits a lower bound around $ \Delta \sigma \simeq 10^{-12} - 10^{-13}$ for most of the modes when $N$ is sufficient large (the precise degree is mode dependent). 
Increasing further the truncation degree often results in a loss of precision for the eigenvalues. 
Similar convergence behaviours have been reported with insulating BC (IN BC) in spheres (see figure 1 in \cite{livermore2005comparison}, or figures 3 and 4 in \cite{li2010optimal}).
This is due to the eigensolver, when the round-off errors in double-precision arithmetic dwarf the truncation errors (which are mode dependent for a given $N$). 
Exponential convergence could be recovered for larger $N$ by using quadruple-precision arithmetic (as considered for another problem in \cite{rieutord2018axisymmetric}). 
Note that the matrices in (\ref{eq:MatrixDynamoPb}) are also found to be severely ill-conditioned for truncation degrees $N\gtrsim20$ in our implementation of the Galerkin method.
This is likely because of the lack of symmetries in the construction of the basis elements, which is unfortunately necessary to account for ellipsoidal geometries in Cartesian coordinates (contrary to previous numerical implementations in spheres\cite{li2010optimal,luo2020optimal}, which can fully exploit separation of variables and the orthogonality of the spherical harmonics). 
Our numerical implementation in double-precision arithmetic is thus currently limited to the large-scale diffusive modes (figure \ref{fig:freedecay}b). 

\begin{figure}
    \centering
    \begin{tabular}{cc}
    \includegraphics[width=0.49\textwidth]{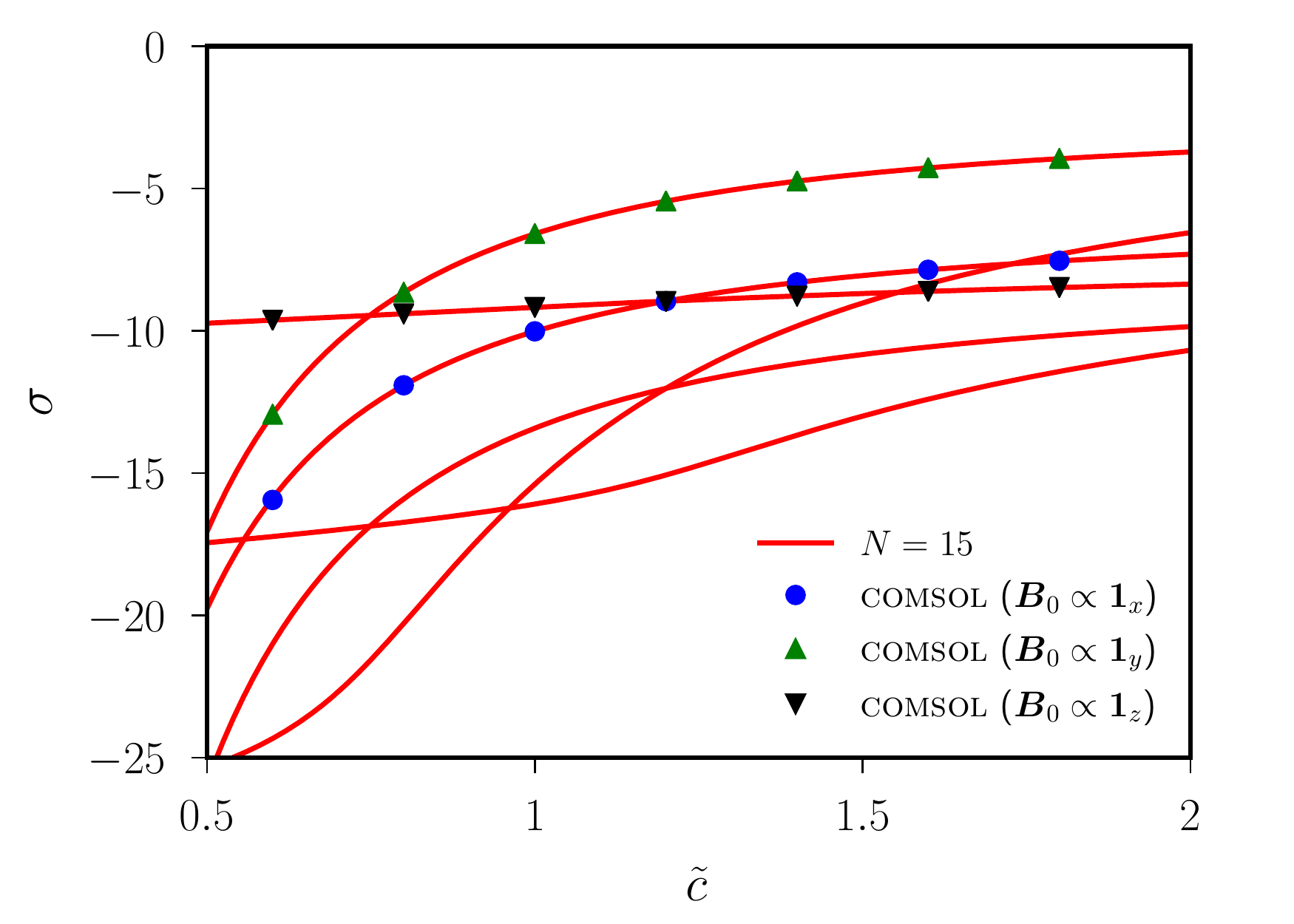} & 
    \includegraphics[width=0.49\textwidth]{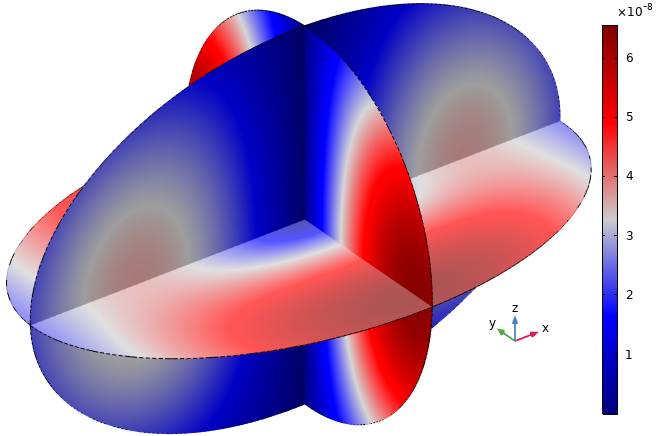} \\
    (a) & (b)
    \end{tabular}
    \caption{(a) Dimensionless decay rate $\sigma$ as a function of polar axis $\tilde{c}$ in triaxial ellipsoids with equatorial ellipticity $\beta=0.44$. Solid curves indicate polynomial solutions at $N=15$, and symbols are the slowest decaying modes obtained with \textsc{comsol} for different initial conditions $\boldsymbol{B}_0$. (b) Three-dimensional rendering of the free-decay magnetic mode for initial condition $\boldsymbol{B}_0 \propto \boldsymbol{1}_z$ (computed with \textsc{comsol}) in a triaxial ellipsoid with $\tilde{c}=0.8$ and $\beta=0.44$.
    Colour bar shows the amplitude of $|\nabla \times \boldsymbol{B}|$. (Online version in colour.)}
    \label{fig:freedecay2}
\end{figure}
 
We can now investigate how the free-decay modes are modified in ellipsoids. 
Analytical results are only available for the axisymmetric toroidal modes in spheroids (because IN BC and PV BC are equal for the axisymmetric toroidal modes in spheroids, see appendix B in \cite{wu2009dynamo}), and so we must rely on numerical calculations to obtain the other modes.
Numerical computations were undertaken using a finite-volume code in \cite{vantieghem2016applications}, finding a relatively good quantitative agreement for the slowest decaying modes (see table 3 in \cite{vantieghem2016applications}). 
We show in figure \ref{fig:freedecay2} a more exhaustive comparison, as a function of the polar axis $\tilde{c}$ in triaxial ellipsoids, between polynomial solutions and high-precision finite-element computations performed with the commercial code \textsc{comsol}.
We have time-stepped the induction equation with \textsc{comsol}, starting from an initial condition $\boldsymbol{B}_0$.
We refer the reader to appendix \ref{appendix:fem} for further numerical details about the finite-element implementation. 
A very good quantitative agreement is found between the two approaches. 
Surprisingly, we also observe that the decay rate depends on the initial field in triaxial ellipsoids. 
Tuning $\boldsymbol{B}_0$ thus allows us to select various free-decay modes in the finite-element computations.
We find three branches for the slowest decaying modes as a function of $\boldsymbol{B}_0$ when $\tilde{c}$ is varied, which coalesce into two branches in spheroids (when $\beta=0$, not shown), whereas a single slowest decaying mode is obtained in spheres.
This effect was overlooked in \cite{vantieghem2016applications}, where the reported decaying mode in the triaxial ellipsoid actually corresponds to the branch with $\boldsymbol{B}_0 \propto \boldsymbol{1}_z$ at $\tilde{c}=1$ in figure \ref{fig:freedecay2}.
We have therefore fully validated our novel spectral method in spherical and triaxial geometries. 
In the following, we set $N= 20$ to have a good frequency (and spatial) convergence for the largest-scale (diffusive) magnetic fields of interest.

\subsection{Illustrative kinematic dynamos}
\begin{figure}
    \centering
    \begin{tabular}{ccc}
    \includegraphics[trim=0 1.5cm 0 1.5cm, clip,width=0.31\textwidth]{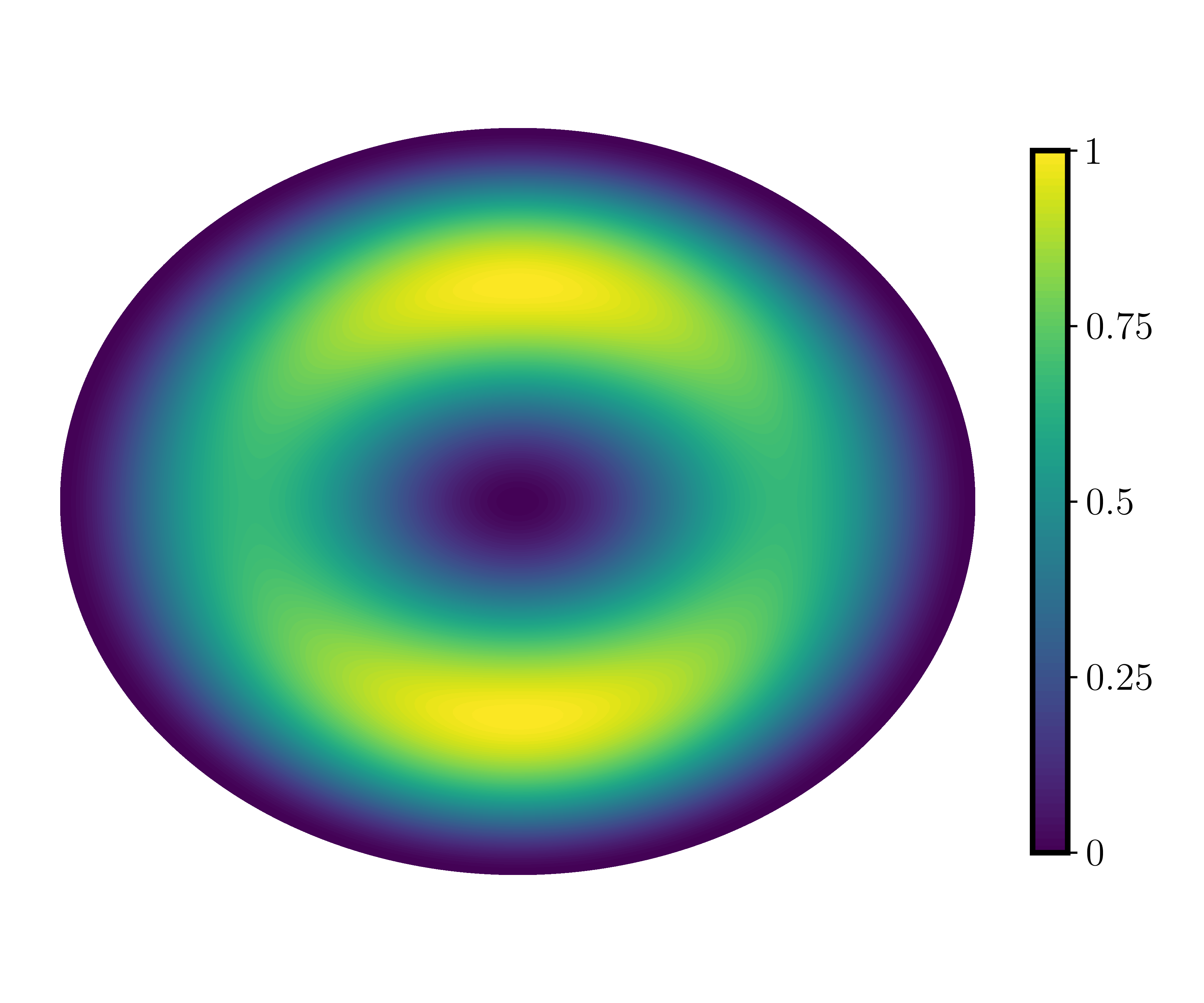} & 
    \includegraphics[trim=0 1.5cm 0 1.5cm, clip,width=0.31\textwidth]{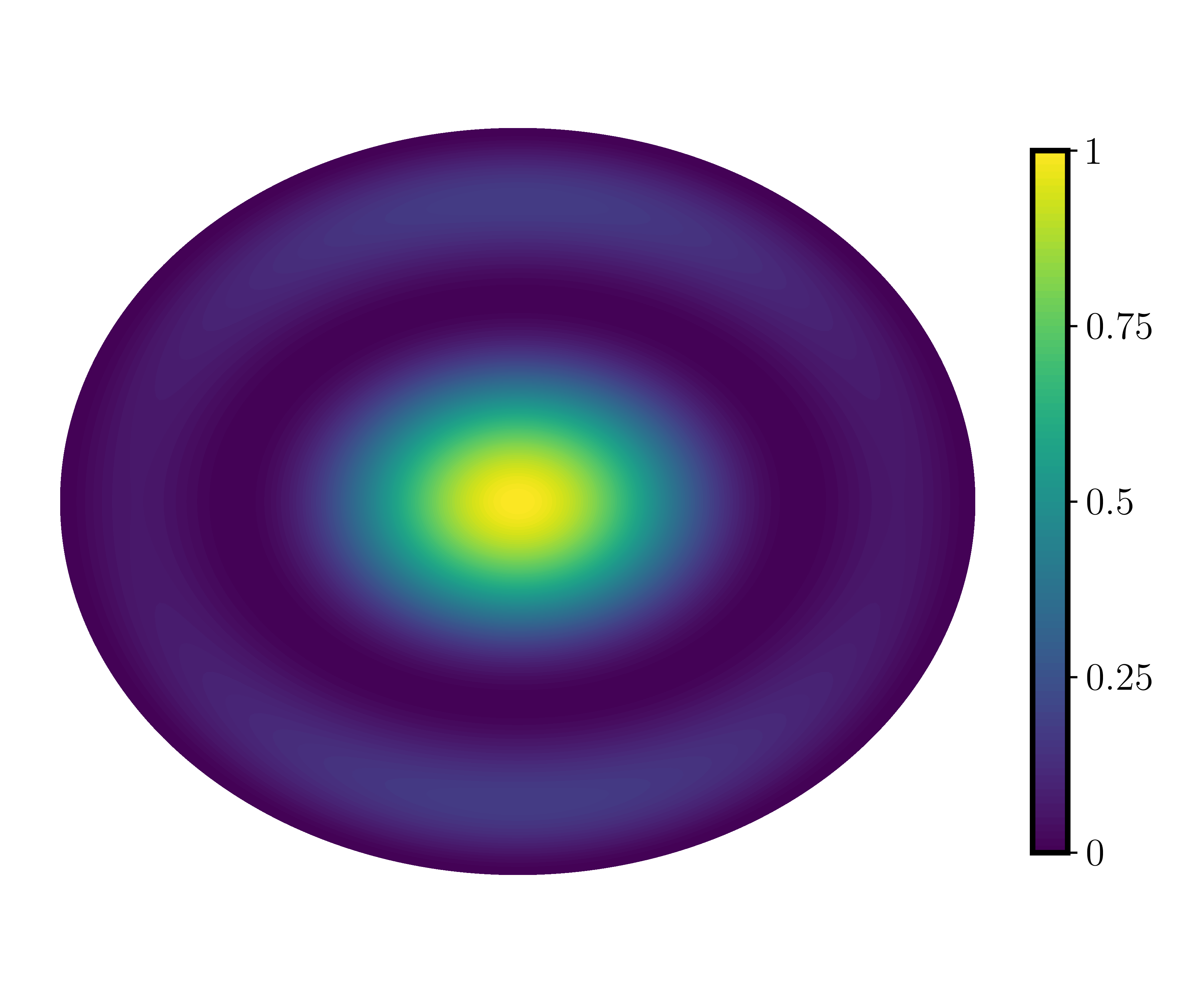} &
    \includegraphics[trim=0 1.5cm 0 1.5cm, clip,width=0.31\textwidth]{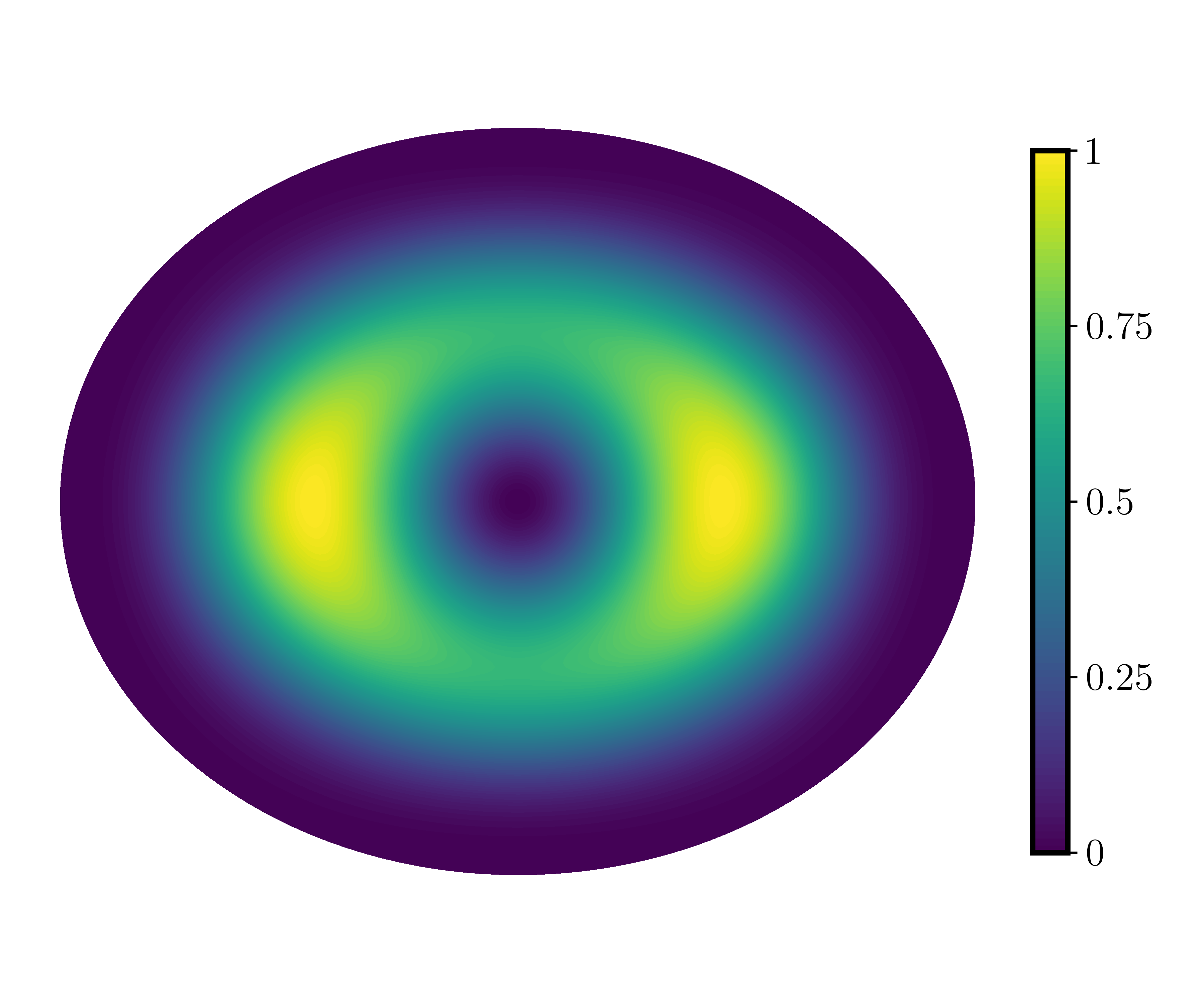} \\
    (a) Flow $T_1^0P_1^0$ with $\epsilon_2 = 0$ & (b) Flow $T_1^0P_1^0$ with $\epsilon_1 = 0$ & (c) Flow $T_1^0P_2^0$ with $\epsilon_1 = 0$ \\
    \end{tabular}
    \caption{Two-dimensional rendering of the (normalised) kinetic energy of toroidal and poloidal components of flows (\ref{eq:flowT10P10P20}) in the plane $z=0$, for an ellipsoid with equatorial ellipticity $\beta=0.2$ and polar axis $\tilde{c}=1$. (Online version in colour.)}
    \label{fig:visuflow}
\end{figure}

We now consider non-vanishing flows $\boldsymbol{v} \neq \boldsymbol{0}$ to explore dynamo action in ellipsoids, which has only received scant attention so far. 
To propose simple dynamo solutions for future benchmark studies, we consider the two large-scale flows defined by
\begin{subequations}
\allowdisplaybreaks
\label{eq:flowT10P10P20}
\begin{align}
    (T_1^0 P_1^0)&:  \quad \boldsymbol{v} = \epsilon_1 \, \nabla \times \left ( [1-F] z \, \boldsymbol{n} \right ) + \epsilon_2 \, \nabla \times \nabla \times ( [1-F]^2 z \, \boldsymbol{n}), \\
    (T_1^0 P_2^0)&: \quad \boldsymbol{v} = \epsilon_1 \, \nabla \times \left ( [1-F] z \, \boldsymbol{n} \right ) + \epsilon_2 \, \nabla \times \nabla \times ( [1-F]^2 [z^2 - (x^2/2) - (y^2/2)] \, \boldsymbol{n} ).
\end{align}
\end{subequations}
The coefficients $[\epsilon_1, \epsilon_2]$ are the dimensionless amplitudes of the toroidal and poloidal components (which control the magnetic Reynolds number). 
The flows are illustrated in figure \ref{fig:visuflow}.
They exactly obey the no-penetration BC $\left . \boldsymbol{v} \boldsymbol{\cdot} \boldsymbol{n} \right|_{\partial V} = 0$, and also the no-slip BC $\left . \boldsymbol{v} \times \boldsymbol{n} \right |_{\partial V} = \boldsymbol{0}$ on the boundary.
Moreover, they reduce to the modified Dudley-James flows in full spheres \cite{livermore2004magnetic,livermore2005comparison,li2010optimal}. 

\begin{figure}
    \centering
    \begin{tabular}{cc}
         \includegraphics[width=0.49\textwidth]{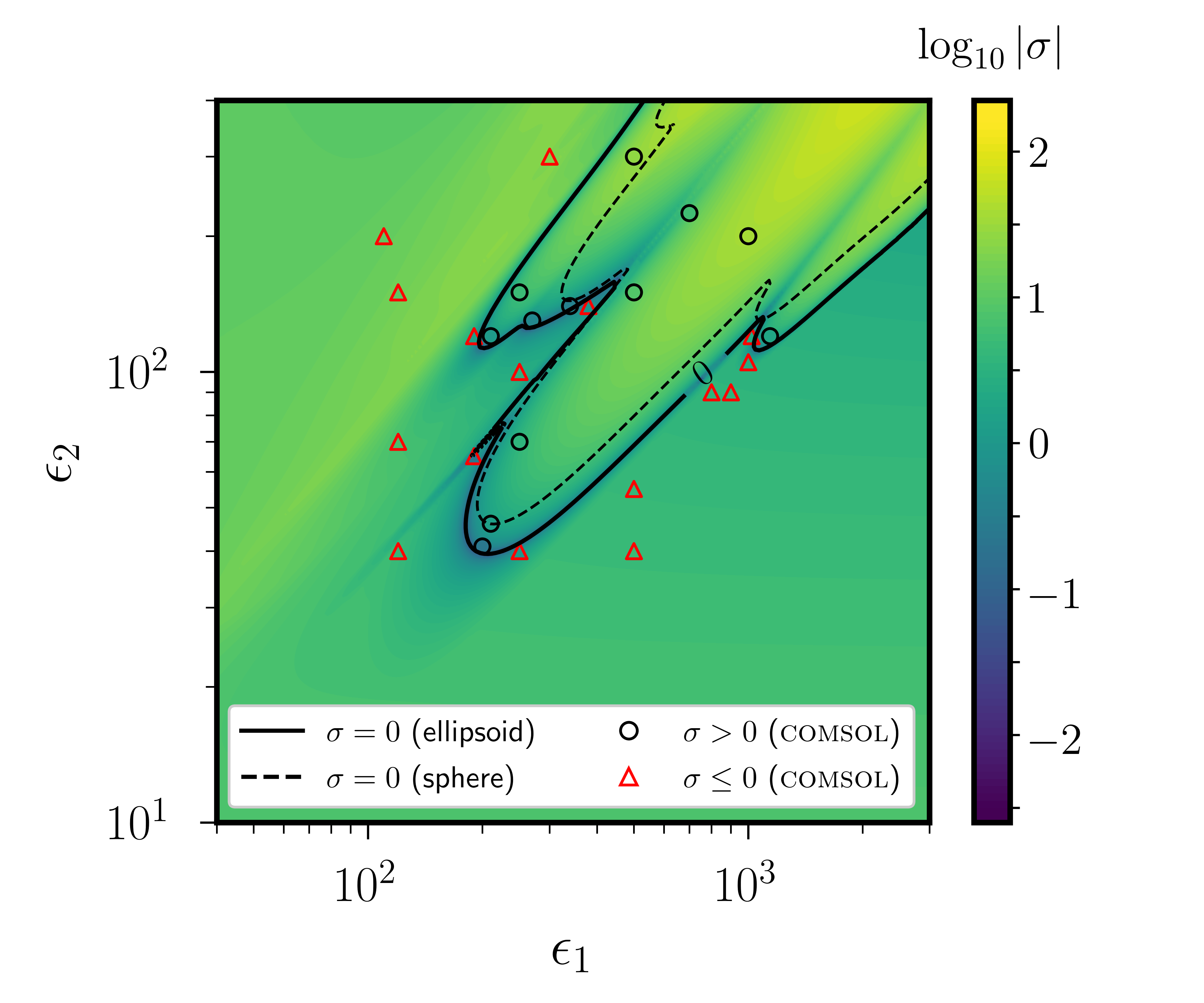} &
         \includegraphics[width=0.49\textwidth]{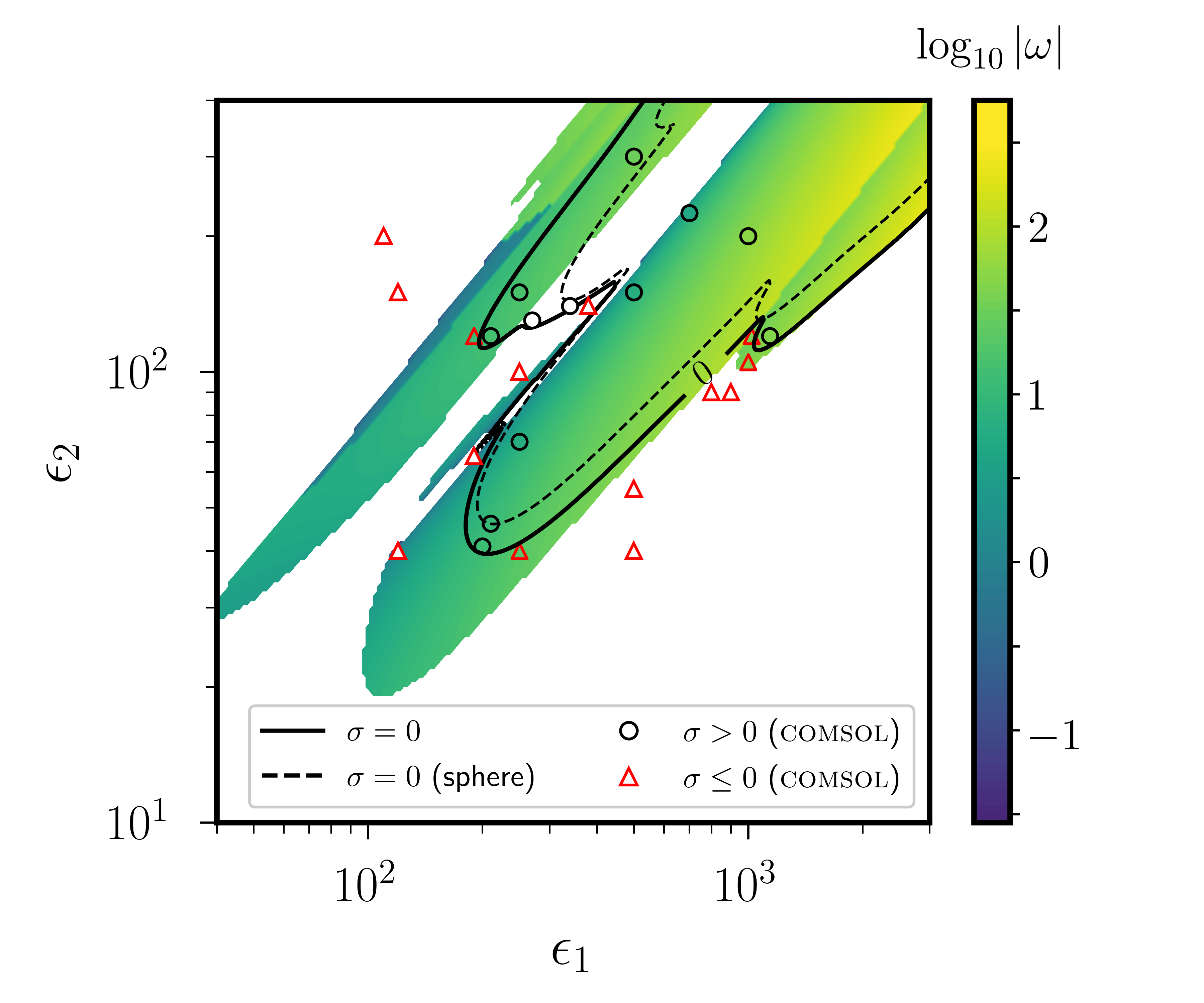} \\[-0.5em]
         \multicolumn{2}{c}{(a) Flow $T_1^0 P_1^0$ with $\tilde{c}=0.95$} \\
         \includegraphics[width=0.49\textwidth]{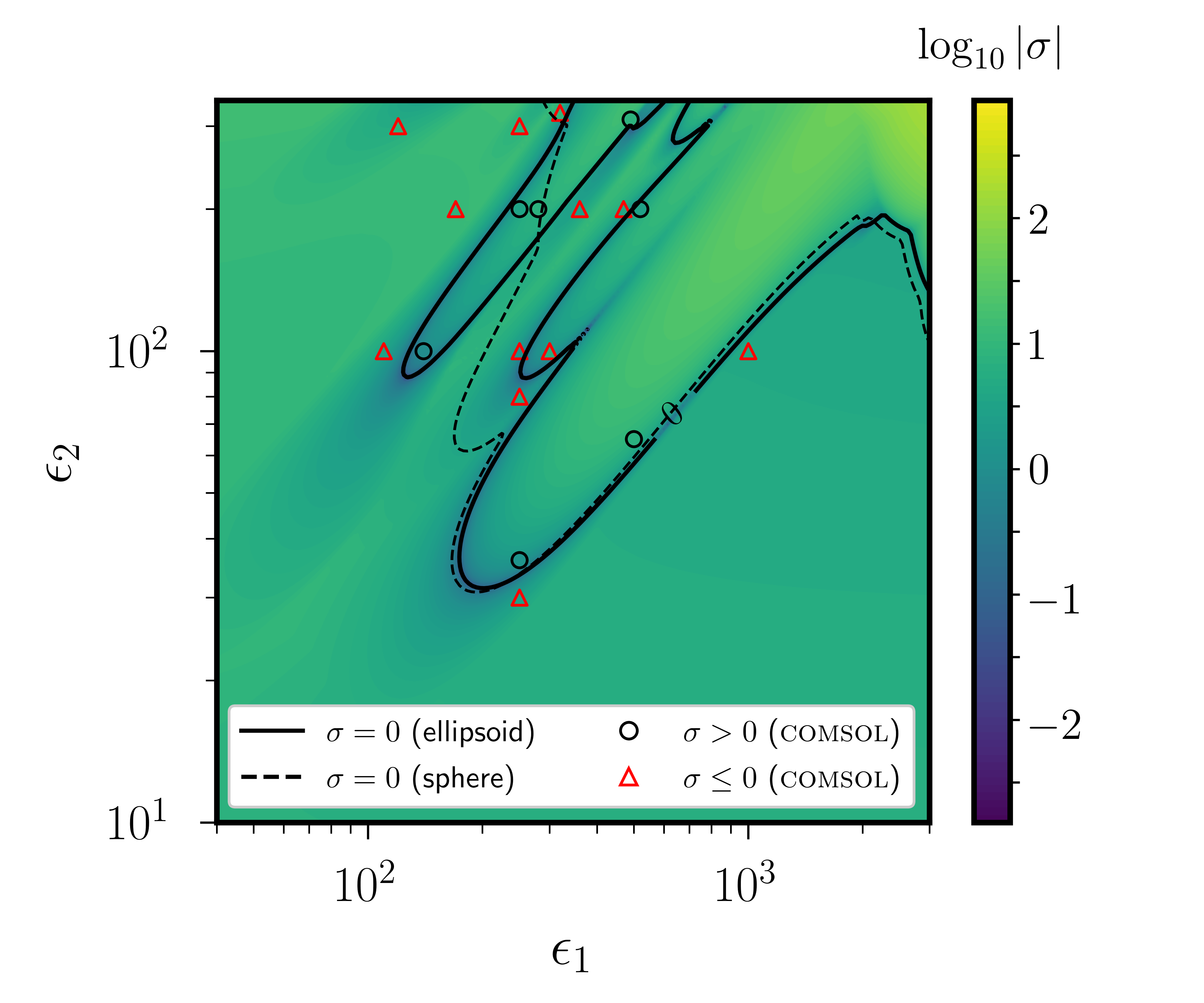} &
         \includegraphics[width=0.49\textwidth]{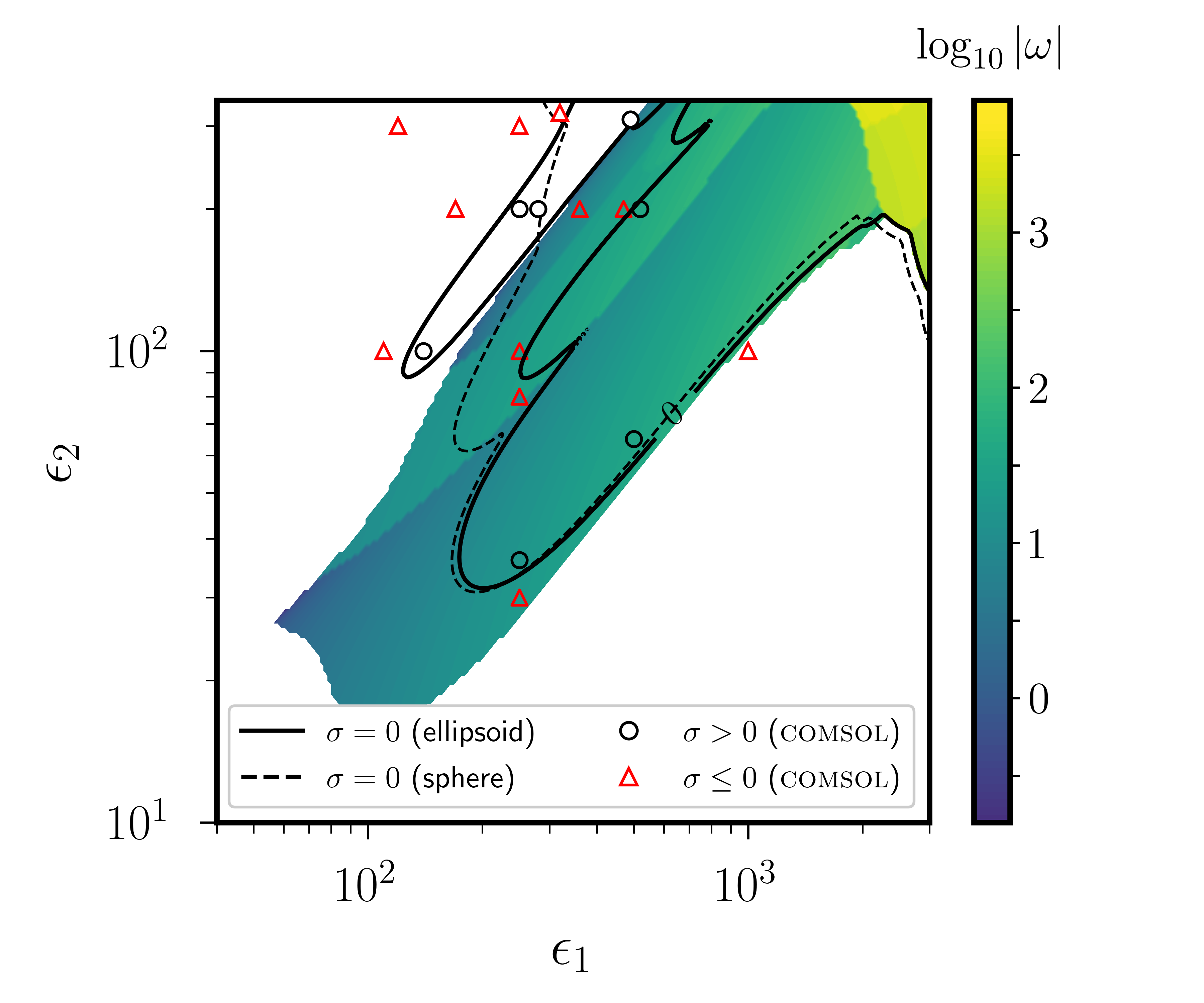} \\[-0.5em]
         \multicolumn{2}{c}{(b) Flow $T_1^0 P_2^0$ with $\tilde{c}=1$} \\
    \end{tabular}
    \caption{Dynamo onset for PV BC as a function of $[\epsilon_1, \epsilon_2]$ for flow $T_1^0 P_1^0$ in panels (a) and flow $T_1^0 P_2^0$ in panels (b), in triaxial ellipsoids with $\beta=0.44$.
    Polynomial calculations at $N=20$ for $200 \times 200$ values of $[\epsilon_1,\epsilon_2]$.
    The marginal state $\sigma=0$ is indicated by the solid black curve for the ellipsoid, and by the dashed black curve for the sphere. 
    Open symbols show \textsc{comsol} results, which agree with the polynomial calculations for the sign of $\sigma$ (the finite size of the symbols can be misleading near the dynamo onset).
    Colour bar shows $\log_{10} |\sigma|$ in the left panels and $\log_{10} |\omega|$ in the right panels (where the white area indicates non-oscillatory solutions $\omega=0$), for both the polynomial and \textsc{comsol} solutions. (Online version in colour.)}
    \label{fig:P10T10T20}
\end{figure}

Solving the kinematic dynamo problem usually amounts to finding the critical value of the magnetic Reynolds number yielding $\sigma = 0$, which is often estimated using root-finding algorithms \cite{livermore2005comparison,li2010optimal} or optimisation methods \cite{chen2018optimal,holdenried2019trio,luo2020optimal}.
Here, we directly compute the dynamo action of the flows in a large region of the parameter space to simplify the comparison between the different numerical methods. 
We show the dynamo onset as a function of $[\epsilon_1, \epsilon_2]$ with PV BC for flow $T_1^0 P_1^0$ in in figure \ref{fig:P10T10T20}a and flow $T_1^0 P_2^0$ in figure \ref{fig:P10T10T20}b, for two ellipsoids with $\beta=0.44$. 
We have chosen this strong ellipticity because the results could be easily reproduced by other non-spherical codes. 
This value is indeed of the order of magnitude of the deformation encountered in most numerical models in ellipsoidal geometries (e.g. \cite{cebron2012magnetohydrodynamic,vantieghem2016applications,reddy2018turbulent}). 
The colour bar in the left panels illustrates $\log_{10} |\sigma|$ of the fastest growing dynamo when $\sigma \geq 0$ (or the slowest decaying mode when $\sigma < 0$), whereas $\log_{10} |\omega|$ is shown in the right panels. 
We have also indicated the results of targeted finite-element computations, which are in perfect agreement for the sign of $\sigma$ (i.e. for the dynamo capability of the flows). 
The values of $\sigma$ and $\omega$ are also in very good quantitative agreement with the polynomial solutions at $N=20$ (relative errors smaller than a few percent are obtained, but they are not observable in the figure given the range of the colour bars). 

Several results are worth commenting on in figure \ref{fig:P10T10T20}. 
Dynamo magnetic fields occur in the form of several tongues of instability within the parameter space, which are surrounded by stable regions. 
There is thus a non-monotonic variation of $\sigma$ with $Rm$ (e.g. as a function of $\epsilon_1$ for a fixed value $\epsilon_2$, or conversely as a function of $\epsilon_2$ for a fixed value $\epsilon_1$). 
Such a succession of unstable-stable tongues is for instance clearly observed in figure \ref{fig:P10T10T20}b when varying $\epsilon_1$ with $10^2 \leq \epsilon_2 \leq 3 \times 10^2$. 
Small variations in $[\epsilon_1,\epsilon_2]$ can so drastically change the dynamo capability of these flows, and this effect certainly exists for other flows with several degrees of freedom. 
Therefore, our results suggest that it could be very difficult to accurately estimate the critical $Rm$ using iterative methods for more complicated flows (due to the presence of local minimums). 
Another striking point is that the nature of the dynamo bifurcation $\sigma=0$ is not the same for the different tongues (as shown in the right panels). 
Dynamo magnetic fields can onset as Hopf bifurcations with $|\omega|\neq0$, or in the form of non-oscillatory modes with $\omega=0$. 
The decaying modes are also modified by the flows, because they can have a non-zero angular frequency in some regions of the parameter space (which contrasts with the non-oscillatory free-decay magnetic modes). 
The angular frequencies can span several orders of magnitude, such that the oscillatory modes are often difficult to accurately compute by integrating the induction equation in time (because of time-step constraints and the limited time of integration). 

\begin{figure}
    \centering
    \begin{tabular}{cc}
         \includegraphics[width=0.49\textwidth]{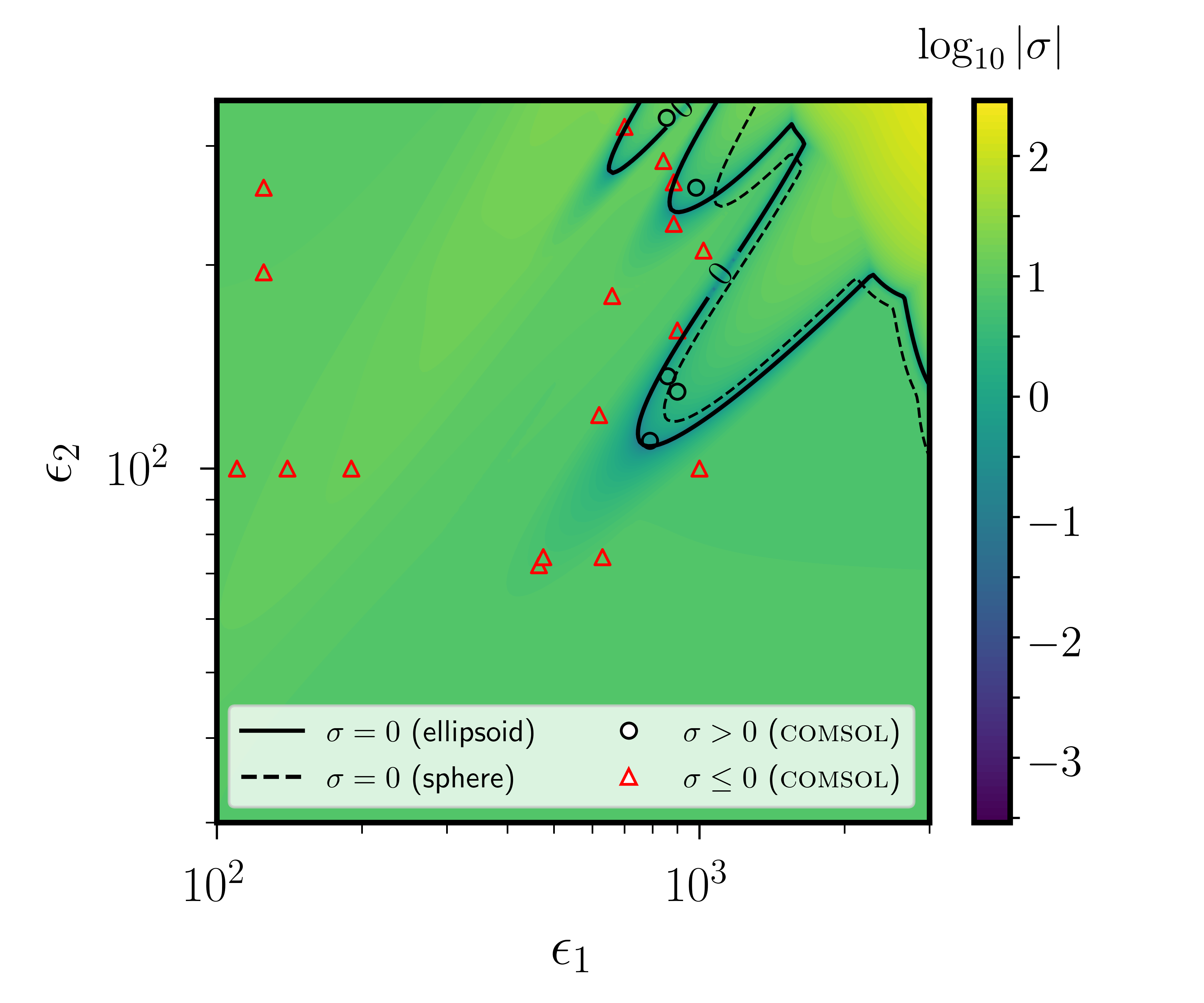} &
         \includegraphics[width=0.49\textwidth]{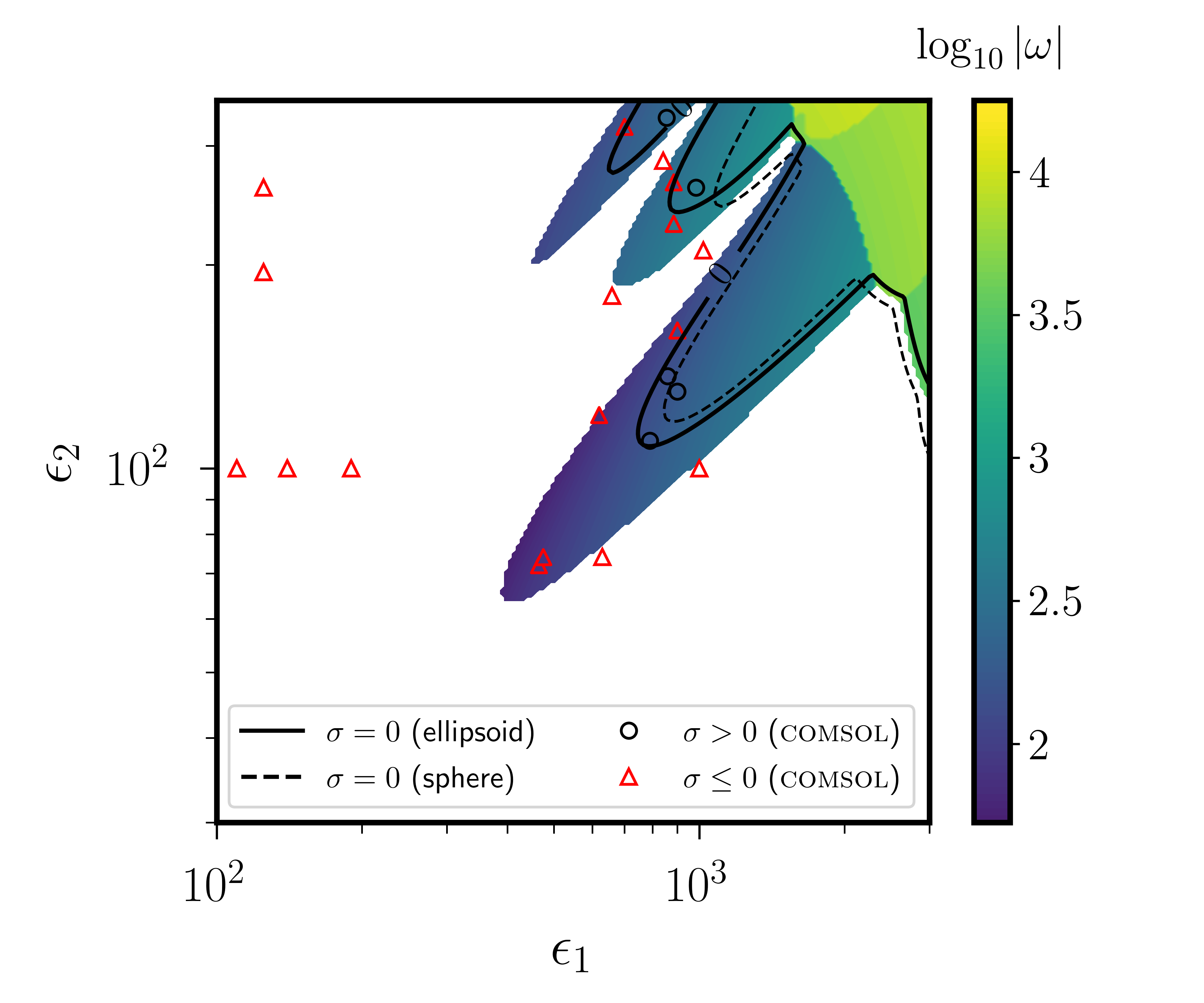} \\
    \end{tabular}
    \caption{Dynamo onset for PC BC as a function of $[\epsilon_1, \epsilon_2]$, for $T_1^0 P_2^0$ flow in a triaxial ellipsoid with $\beta=0.44$ and $\tilde{c}=1$.
    Polynomial calculations at $N=20$ for $200 \times 200$ values of $[\epsilon_1,\epsilon_2]$.
    The marginal state $\sigma=0$ is indicated by the solid black curve for the ellipsoid, and by the dashed black curve for the sphere. 
    Open symbols show \textsc{comsol} results, which agree with the polynomial calculations for the sign of $\sigma$ (the finite size of the symbols can be misleading near the dynamo onset).
    Colour bar shows $\log_{10} |\sigma|$ in the left panels and $\log_{10} |\omega|$ in the right panels (where the white area indicates non-oscillatory solutions $\omega=0$), for both the polynomial and \textsc{comsol} solutions. (Online version in colour.)}
    \label{fig:T10P20supra}
\end{figure}

We can now compare how using PC BC affects the dynamo onset computed with PV BC.
We find that the numerical eigenvalues $\lambda$ are (almost) identical for PC BC and PV BC when considering flow $T_1^0 P_1^0$ in ellipsoids (not shown here, see below),
although the magnetic field morphology is completely different. 
This certainly results from underlying mathematical symmetries of the flow $T_1^0 P_1^0$ in ellipsoidal geometries \cite{favier2013growth}. 
The other flow $T_1^0 P_2^0$ does not however exhibit such a property, as illustrated in figure \ref{fig:T10P20supra}.
Polynomial solutions at $N=20$ are again in excellent quantitative agreement with the finite-element computations, demonstrating the robustness of the results.
We observe several tongues of instability that all correspond here to Hopf bifurcations (at least in the illustrated parameter space), and the dynamo fields have much larger angular frequencies $\omega$ than with PV BC.
More importantly, we observe that using PC BC strongly weakens the dynamo capability of the flow. 
Quantitatively, this is evidenced by computing the smallest value $Rm_c \simeq 50$ that characterises the marginal state $\sigma=0$ for PV BC in figure \ref{fig:P10T10T20}b (at the bottom edge of the central tongue), whereas we obtain the smallest critical value $Rm_c \simeq 200$ at the edge of the lower tongue for PC BC in figure \ref{fig:T10P20supra}. 
The observed differences between the two BC are stronger than previously reported for other prescribed flows in spheres (see figure 2 in \cite{favier2013growth}). 
This is also significantly different from the situation explored in \cite{luo2020optimal}, where the optimised flows were almost identical for the two BC and yielded nearly the same smallest critical value for $Rm$. 
Our results suggest instead that, for a given flow, the dynamo onset can be strongly affected by using PV BC or PC BC. 

Finally, it is worth assessing how the dynamo onset evolves with the ellipsoidal deformation. 
We have thus also shown in figures \ref{fig:P10T10T20} and \ref{fig:T10P20supra} the onset curves $\sigma=0$ for the sphere, which are expected to represent the low-ellipticity regime $\beta \to 0$ for these particular flows (which continuously vary from the sphere to the ellipsoid). 
The main unstable tongue is found to be weakly sensitive to the ellipsoidal geometry, such that the smallest critical Reynolds number weakly depends on the ellipsoidal geometry for these two flows.
However, the other unstable tongues strongly vary with $\beta$. 
The ellipticity can either enhance the dynamo capability of the two considered flows or weaken dynamo action (e.g. see the intermediate tongue in figure \ref{fig:P10T10T20}b). 
This simple example outlines that extrapolating dynamo results from spheres to ellipsoids (or vice versa) is not straightforward, even for such simple flows.

\section{Discussion}
\label{sec:discussion}
We have shown that PV BC or PC BC can lead to very different estimates for the dynamo onset in ellipsoids (depending on the considered flows). 
Another major difference between these two BC is that the magnetic field is trapped in the fluid volume with PC BC. 
PV BC are thus in principle more suitable than PC BC for planetary dynamo models, because planetary magnetic fields are allowed to escape the fluid system to match an exterior potential field (when the exterior is modelled by an electrical insulator). 
PV BC are indeed often believed to yield qualitatively similar results to IN BC (e.g. for convective dynamos in plane-layer \cite{thelen2000dynamo} or spherical \cite{kageyama1997velocity,harder2005finite} geometries), but the two BC must be further compared as the similarity could be flow dependent. 

\begin{figure}
    \centering
    \begin{tabular}{cc}
        \includegraphics[width=0.49\textwidth]{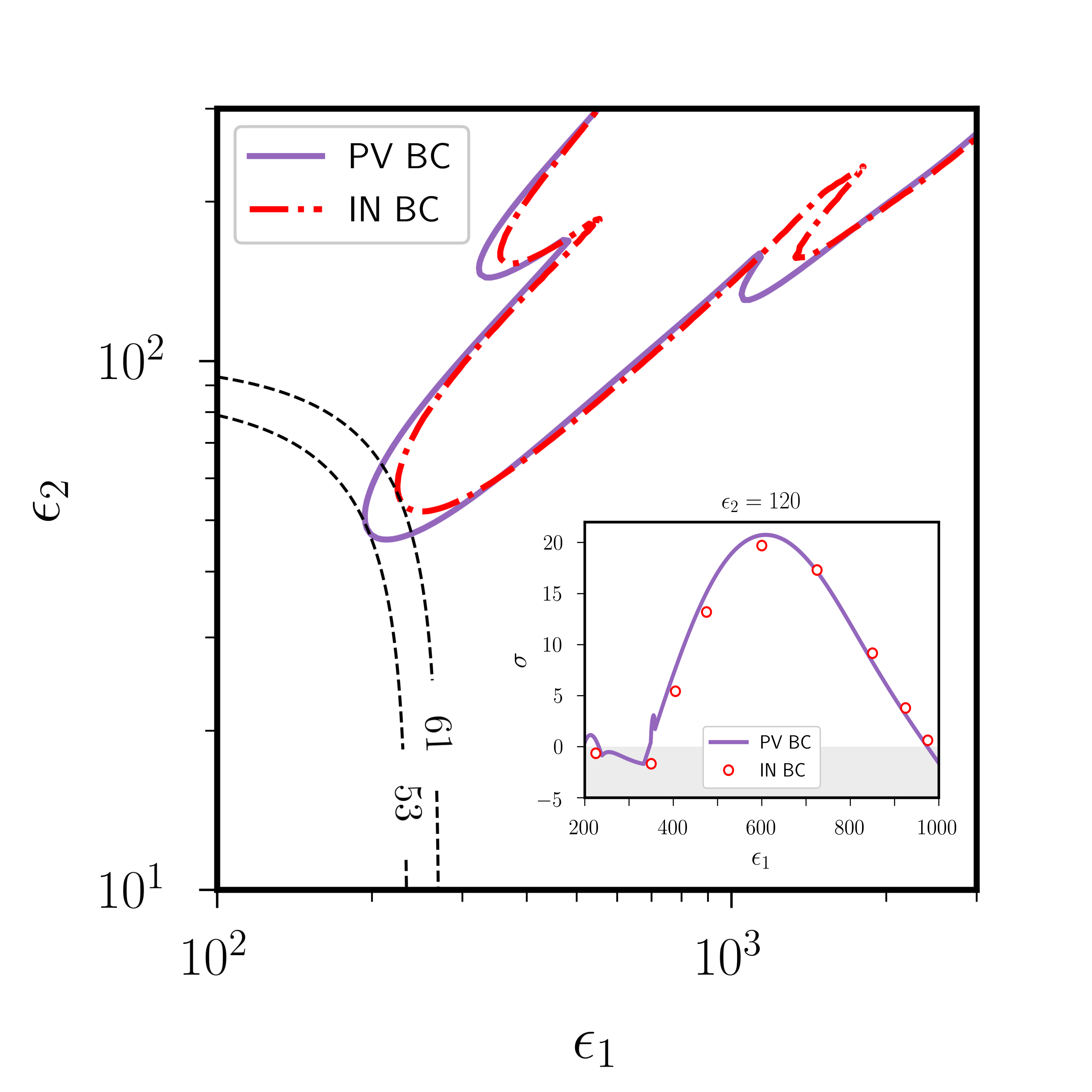} &
        \includegraphics[width=0.49\textwidth]{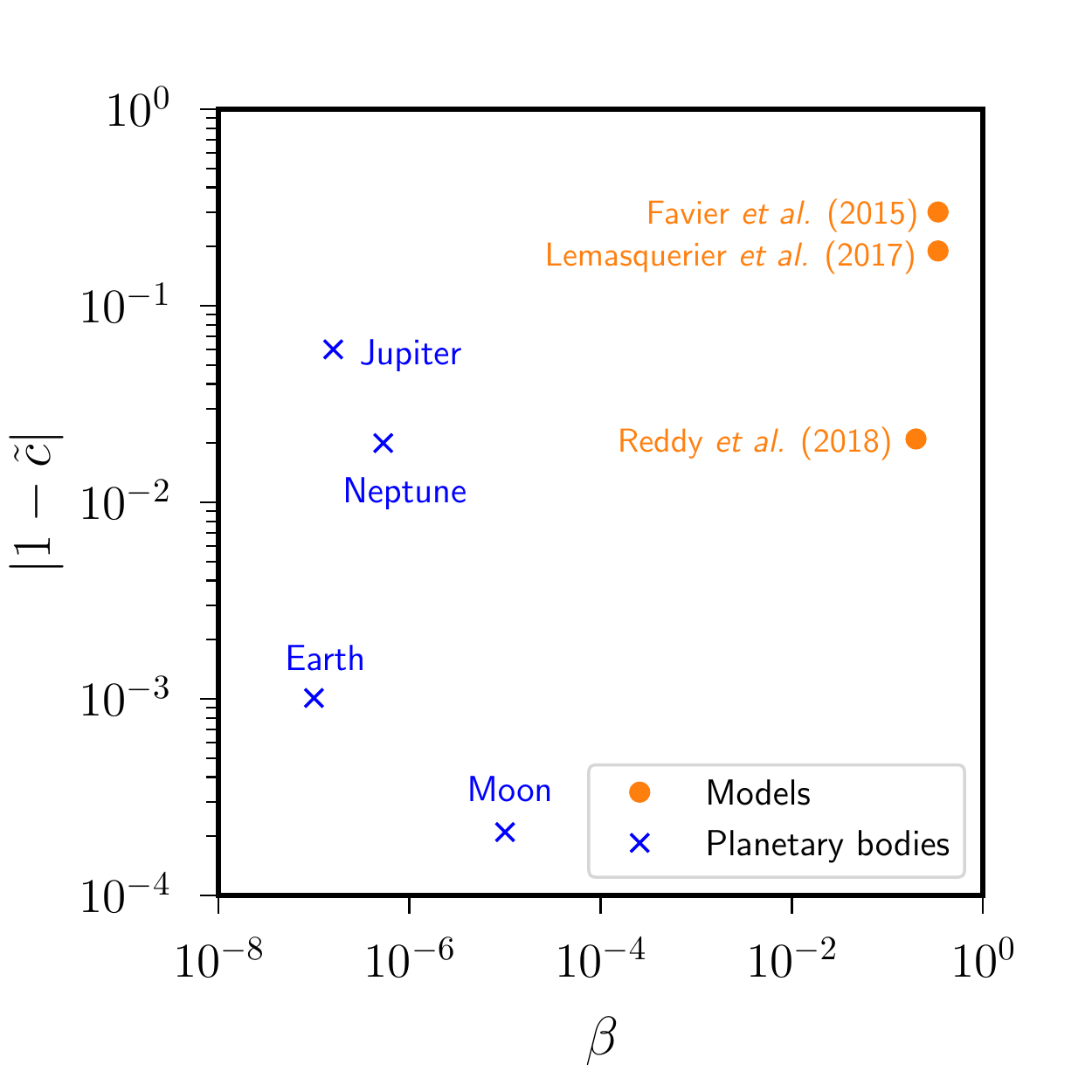} \\
        (a) & (b) \\
    \end{tabular}
    \caption{Comparison between PV BC, and IN BC for dynamo onset with flow $T_1^0 P_1^0$. 
    (a) Marginal curve $\sigma=0$ in spheres for polynomial calculations at $N=20$. The solid purple curve shows PV BC, and the dash-dot red curve IN BC (see the electronic supplementary material for the corresponding polynomial description).
    The thin black dashed curves indicate values of $Rm$ given by formula (\ref{eq:RmU}). 
    Inset shows $\sigma$ as a function of $\epsilon_1$ with $\epsilon_2=120$ in a triaxial ellipsoid with $\beta=0.44$ and $\tilde{c}=0.95$, for PV BC (purple solid curve) and IN BC (red empty circles, \textsc{comsol} solutions). 
    (b) Typical values of dimensionless polar axis $\tilde{c}$ and equatorial ellipticity $\beta$ for some numerical and experimental flow models in ellipsoids \cite{favier2015generation,lemasquerier2017libration,reddy2018turbulent}, and for a few planetary bodies in the Solar System (values taken from table 2 in \cite{wicht2010theory}). (Online version in colour.)}
    \label{fig:insulating1}
\end{figure}

We can undertake such a comparison for the two flows $T_1^0 P_1^0$ and $T_1^0 P_2^0$ in a full sphere, because polynomial expansions for the magnetic field satisfying IN BC can be readily obtained in this geometry (see the discussion in the electronic supplementary material).
The resulting dynamo problem can then be solved using a projection method as presented in \S\ref{sec:numerics}, except that the diffusion operator must be written in the non-symmetric form $\boldsymbol{D}_{ij} = -\langle \boldsymbol{e}_i, \nabla^2 \boldsymbol{e}_j \rangle$. 
We show in figure \ref{fig:insulating1}a the comparison between PV BC and IN BC for flow $T_1^0 P_1^0$ in a full sphere. 
We obtain qualitatively and quantitatively very similar results for the dynamo onset $\sigma=0$. 
In particular, the smallest critical Reynolds number is only weakly reduced from $Rm_c\simeq61$ for IN BC to $Rm_c\simeq53$ for PV BC. 
A similar observation can be made for flow $T_1^0 P_2^0$, with $Rm_c\simeq55$ for IN BC \cite{livermore2004magnetic} and $Rm_c\simeq 43$ for PV BC (not shown). 
It is also worth comparing the two BC in ellipsoidal geometries but, unfortunately, IN BC cannot be implemented as easily as PV BC in ellipsoidal geometries.
One should indeed employ ellipsoidal harmonic expansions to do so (owing to the global nature of IN BC), but the ellipsoidal harmonics do not admit simple explicit Cartesian expansions for numerical computations \cite{dassios2012ellipsoidal}. 
Other numerical strategies have been proposed to explore the effect of IN BC in non-spherical geometries (e.g. using the boundary-element method \cite{iskakov2004integro}, or non-orthogonal coordinates \cite{ivers2017kinematic}). 
Here, we can compare the two BC in ellipsoids by accurately approximating IN BC using the finite-element method (see details in appendix \ref{appendix:fem}). 
A good agreement between PV BC and IN BC is found for the dynamo growth rate for flow $T_1^0 P_1^0$ when $\sigma$ is beyond the dynamo onset (see the inset in figure \ref{fig:insulating1}a). 

\begin{figure}
    \centering
    \begin{tabular}{cc}
        \includegraphics[width=0.49\textwidth]{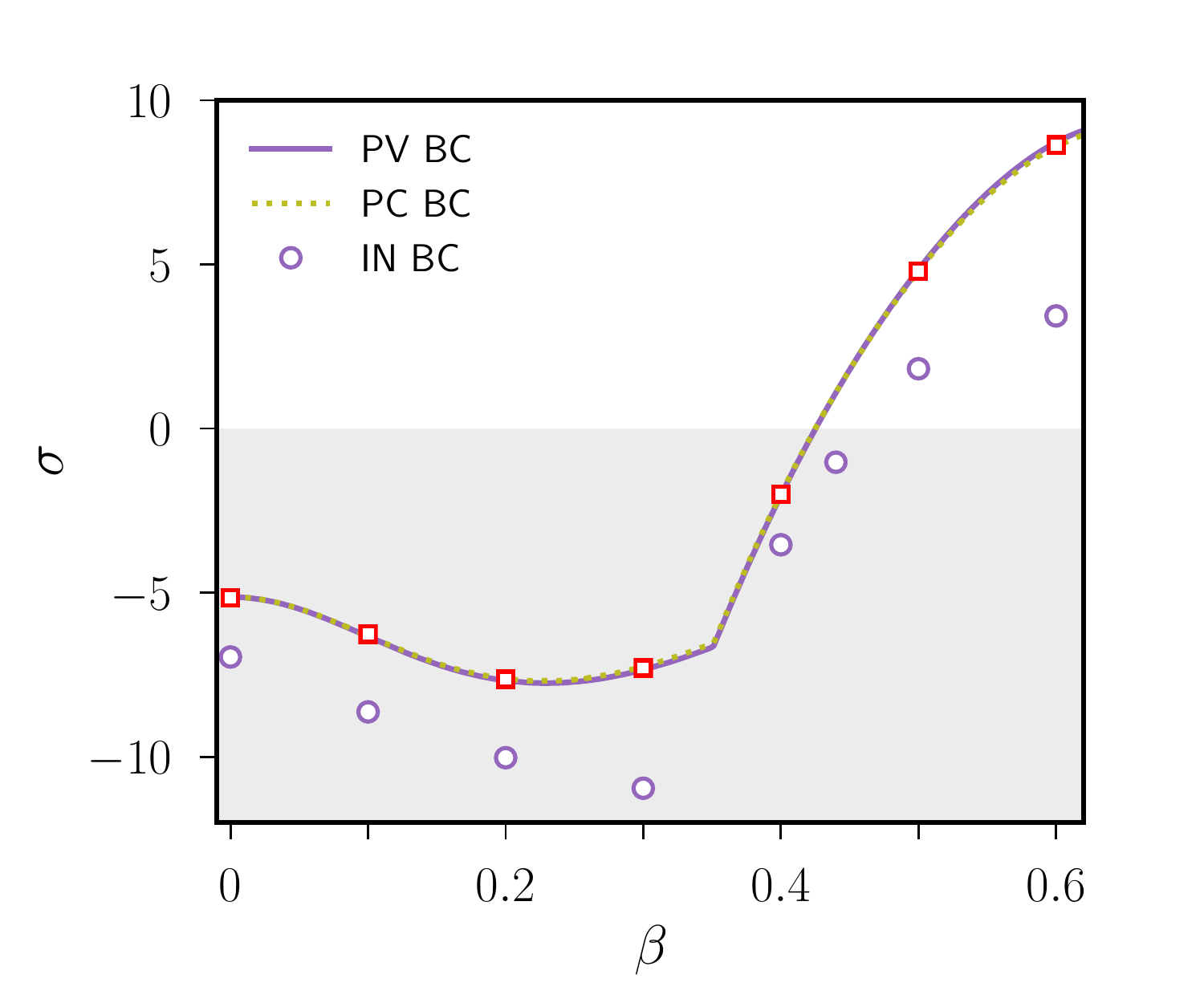} &
        \includegraphics[width=0.49\textwidth]{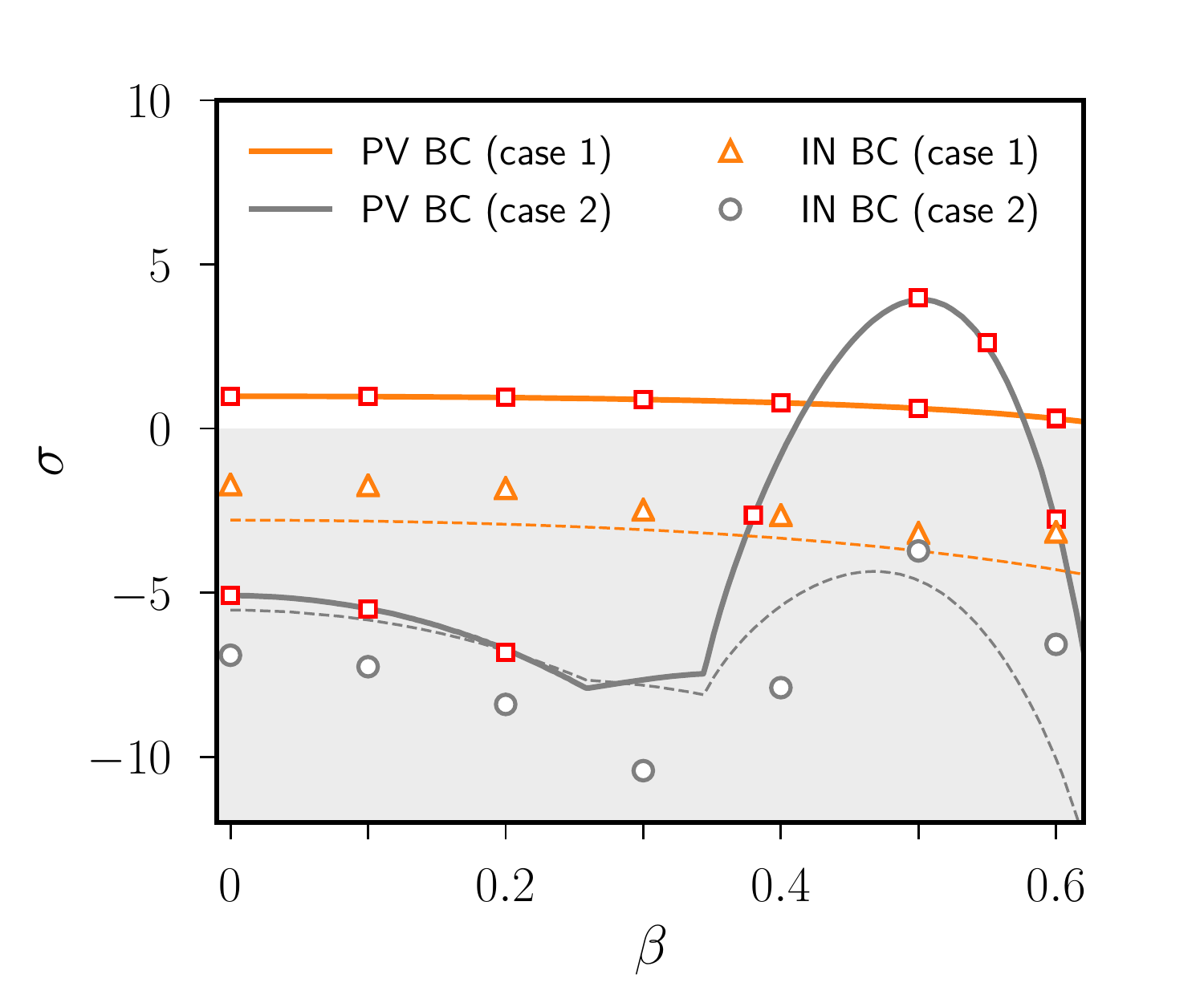} \\
        (a) Flow $T_1^0 P_1^0$ with $\tilde{c}=0.95$ & (b) Flow $T_1^0 P_2^0$ with $\tilde{c}=1$ \\
    \end{tabular}
    \caption{Evolution of $\sigma$ as a function of $\beta$ in triaxial ellipsoids, for flow $T_1^0 P_1^0$ in (a) and flow $T_1^0 P_2^0$ in (b).
    Comparison between IN BC (open symbols, computed with \textsc{comsol}) and PV BC (coloured curves, polynomial solutions at $N=20$).
    (a) Fixed values $\epsilon_1=210$  and $\epsilon_2=120$.
    (b) Case 1: fixed values $\epsilon_1=190$  and $\epsilon_2=35$. Case 2: fixed values $\epsilon_1=165$ and $\epsilon_2=120$. 
    In the two panels, red open squares indicate \textsc{comsol} computations with PV BC.
    Thin coloured dashed curves show the arithmetic average between PV BC and PC BC. (Online version in colour.)}
    \label{fig:profilbeta}
\end{figure}

\begin{figure}
    \centering
    \begin{tabular}{cc}
        \includegraphics[width=0.45\textwidth]{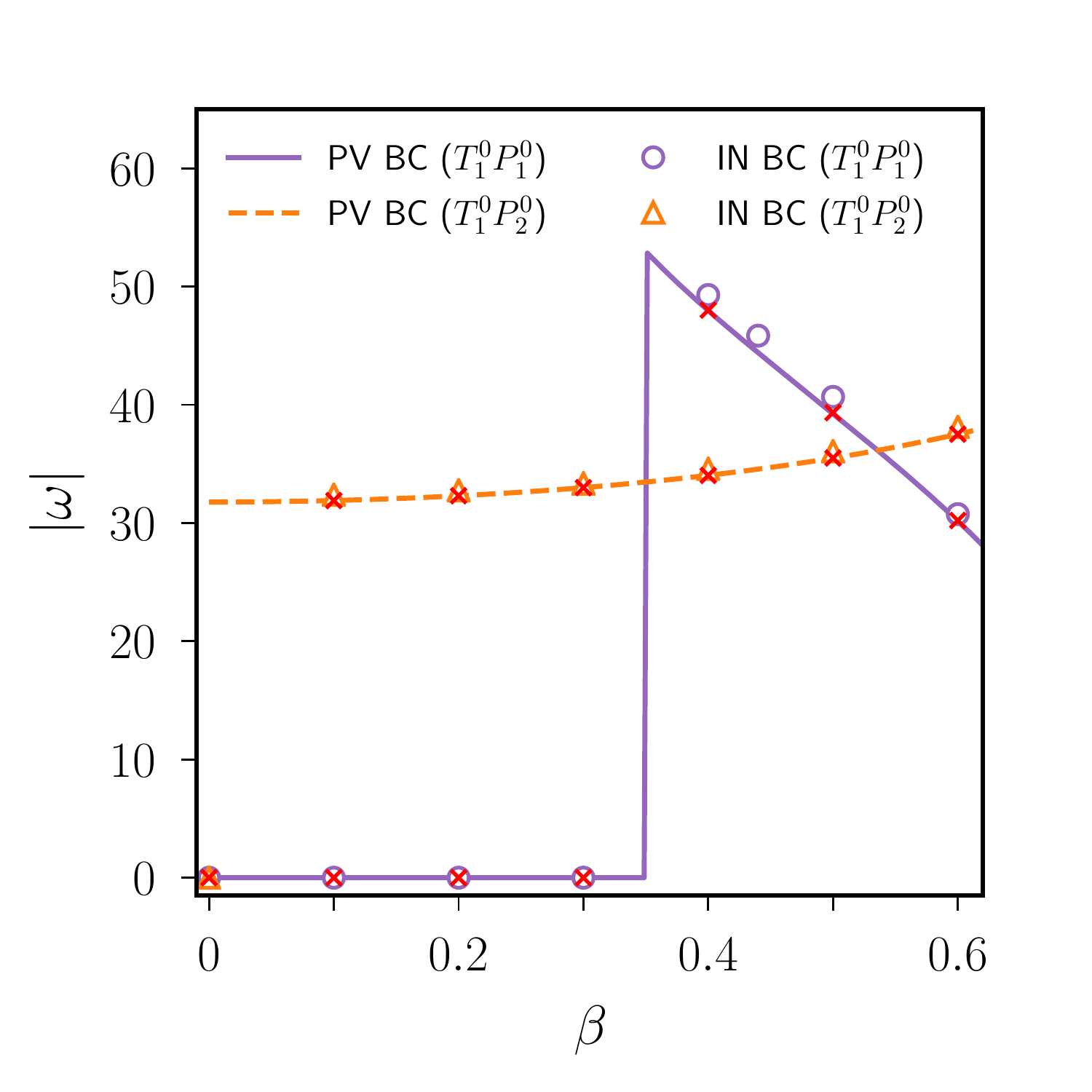} &
        \includegraphics[trim=0 0 0 0,clip,width=0.47\textwidth]{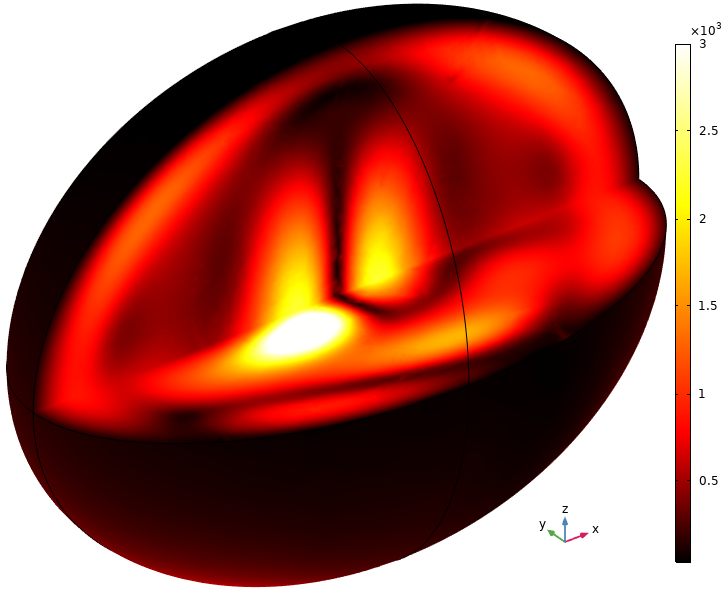} \\
        (a) & (b) \\
    \end{tabular}
    \caption{(a) Dimensionless angular frequency $|\omega|$ as a function of $\beta$ in triaxial ellipsoids with PV BC and IN BC. Flow $T_1^0 P_1^0$ with $\tilde{c}=0.95$, $\epsilon_1=210$ and $\epsilon_2=120$ (as in figure \ref{fig:profilbeta}a). 
    Flow $T_1^0 P_2^0$ with $\tilde{c}=1$, $\epsilon_1=190$ and $\epsilon_2=35$ (case 1 in figure \ref{fig:profilbeta}b).
    Red crosses indicate \textsc{comsol} simulations with PV BC.
    (b) Three-dimensional rendering of $||\boldsymbol{B}||$ for a dynamo magnetic field with IN BC for flow $T_1^0 P_1^0$ in a triaxial ellipsoid with $\tilde{c}=0.95$ and $\beta=0.5$ (as in figure \ref{fig:profilbeta}a). (Online version in colour.)}
    \label{fig:visuinsulating}
\end{figure}

Other ellipsoidal geometries also have to be considered, since the dynamo onset is certainly ellipsoid dependent (e.g. as shown for PV BC in figure \ref{fig:P10T10T20}).
Before doing so, it is important to estimate the typical amplitude of deformation encountered in planetary bodies on the one hand, and that considered in numerical (or experimental) models on the other hand. 
As illustrated in figure \ref{fig:insulating1}b, planetary bodies are weakly deformed (with $10^{-8} \leq \beta \leq 10^{-5}$, $10^{-2} \leq |1-\tilde{c}| \leq 10^{-1}$ for gaseous planets, and $|1-\tilde{c}| \leq 10^{-3}$ for telluric planets), whereas the models usually account for much larger deformations (i.e. $0.7 \leq \tilde{c} \leq 1.1$, and $\beta \geq 10^{-1}$ for non axisymmetric geometries).
Consequently, to be useful for planetary extrapolations, PV BC and IN BC should give similar results in both strongly deformed and weakly deformed ellipsoids. 
The evolution of $\sigma$ as a function of $\beta$ is illustrated in figure \ref{fig:profilbeta} for the two flows. 
A broad quantitative agreement between the two BC is observed for flow $T_1^0 P_1^0$ but, quantitatively, $\sigma$ is slightly overestimated by using PV BC. 
The discrepancies are more pronounced for flow $T_1^0 P_2^0$, even if the trend qualitatively agrees between the two BC.
This could be particularly awkward close to the dynamo onset, since PV BC can predict unstable dynamo magnetic fields that are instead decaying when considering IN BC (see the unstable tongue around $\beta=0.5$). 
A better quantitative agreement with IN BC is sometimes found by taking the arithmetic average between PV BC and PC BC (figure \ref{fig:profilbeta}b). 
This might result from the fact that IN BC are somehow intermediate between PV BC and PC BC (the magnetic field could be seen as a combination between purely radial and tangential fields at an insulating boundary). 
We also illustrate in figure \ref{fig:visuinsulating}a the comparison for the angular frequency of the magnetic field, showing that IN BC and PV BC are also in good agreement for $\omega$. 
We finally illustrate in figure \ref{fig:visuinsulating}b a typical dynamo magnetic field obtained with IN BC. 
The typical scale of the field is comparable to the length scale of the flow, and we can observe a low magnetic field near the boundary (which can be expected since the flow is no-slip on the boundary).

Therefore, our results suggest that PV BC could reasonably approximate the dynamo growth rate with IN BC in ellipsoids when sufficiently far from the dynamo threshold (since PV BC tend to overestimate $\sigma$). 
We can now come back to our physical motivation, which is the study of mechanically driven dynamos in ellipsoidal planetary cores. 
We have shown above that the ellipsoidal deformation does not affect the dynamo capability of the flow in a simple way, confirming the complexity of dynamo action in ellipsoidal geometries.
The ellipsoidal deformation can indeed either enhance or weaken dynamo action in moderately deformed ellipsoids (as observed for the different BC in figures \ref{fig:P10T10T20}, \ref{fig:T10P20supra} and \ref{fig:profilbeta}). 
It could also lead to turbulent flows in planetary bodies subject to mechanical forcings \cite{le2015flows}, possibly sustaining saturated dynamos in non-spherical geometries. 
Considering a purely spherical domain would instead lead to mechanically driven dynamos only tied to Ekman pumping \cite{tilgner2005precession}, or to internal shear layers \cite{lin2016precession} whose effects may be negligible for planetary applications \cite{cebron2019precessing}. 
Ellipsoidal geometries should thus be modelled in dynamo studies, but results in strongly deformed ellipsoids should always be interpreted with caution for planetary applications. 

\section{Conclusions and future works}
\label{sec:ccl}
We have presented in this work a polynomial spectral method to solve the dynamo problem in full ellipsoids, which are directly relevant for some planetary bodies. 
We have thoroughly assessed the accuracy of the novel method against analytical predictions in spheres, and targeted finite-element computations in ellipsoids. 
We have first computed the free-decay modes of the induction equation (correcting previous numerical findings in ellipsoids \cite{vantieghem2016applications}), and then explored the dynamo capability of some simple velocity fields.  
Our results could be used as accurate dynamo benchmarks for future studies in ellipsoidal geometries. 
We have finally compared the effects of the magnetic BC, showing that PV BC could broadly approximate IN BC (since they tend to overestimate the dynamo growth rate). 

This work could be extended by considering more realistic velocity fields within the kinematic approximation, such as time-dependent velocity fields that can favour dynamo action (owing to the non-normality of the induction equation \cite{tilgner2008dynamo}).
A prerequisite would be to implement extended-precision arithmetic, to be able to accurately compute smaller-scale solutions with polynomial degrees $N \geq 20$. 
Then, inertial modes \cite{vantieghem2014inertial} would particularly deserve consideration, since they are ubiquitous in rapidly rotating fluids \cite{zhang2017theory} and may enhance dynamo action \cite{moffatt1970dynamo} (contrary to an isolated inertial mode \cite{herreman2011stokes}). 
Floquet theory may be used to compute the dynamo growth rate for time periodic flows, but solving the corresponding numerical problem could be very difficult.
The dimension of the polynomial space indeed grows rapidly with the polynomial degree $N$ as $\mathcal{O}(N^3)$.
Since Floquet theory would require time-stepping $\mathcal{O}(N^6)$ coefficients over one period, the required computational power would certainly become too demanding at high resolution.
Directly time-stepping the induction equation could thus be more affordable to explore the dynamo capability of prescribed periodic flows (and could also account for non-periodic flows). 
Transient dynamo growth could also be explored, since subcritical growth has been reported in spheres for simple steady flows \cite{livermore2006transient}. 
Next, instead of imposing the velocity field, it would be worth searching for the optimal spatial structure of the flow (among all permissible polynomial fields) yielding the most efficient dynamo magnetic field. 
We could indeed extend previous dynamo variational algorithms in spheres \cite{chen2018optimal,holdenried2019trio,luo2020optimal} to the ellipsoid, to determine the minimum magnetic Reynolds number for dynamo action in ellipsoidal geometries.
Simulating saturated dynamo fields in ellipsoids is finally a long-term endeavour, but it requires an efficient algorithm to be found for the nonlinear terms.

\enlargethispage{20pt}

\dataccess{Additional data are provided in the electronic supplementary material. The source code \textsc{shine} is available at \url{https://bitbucket.org/vidalje/shine/download}.}

\aucontribute{This work is an original idea of J.V., who designed the study and implemented the Galerkin algorithm. D.C. conducted the finite-element computations. J.V. and D.C. discussed and approved the methods and results presented in the article. J.V. drafted the paper, and both authors gave final approval for submission.}

\competing{The authors declare that they have no competing interests.}

\funding{This work received funding from the European Research Council (ERC) under the European Union's Horizon 2020 research and innovation programme (grant agreement No 847433, \textsc{theia} project).}

\ack{We acknowledge the four anonymous referees, whose valuable comments helped us to improve the clarity and quality of the manuscript.}

\appendix
\numberwithin{equation}{section}

\section{Finite-element computations}
\label{appendix:fem}
To carefully validate our polynomial spectral method accounting for local BC, we solve induction equation (\ref{eq:inductionA}) for the magnetic potential $\boldsymbol{A}$ with standard finite-element computations (using the commercial software \textsc{comsol}). 
The fluid volume is discretised with an unstructured mesh made of tetrahedral finite elements, and we use cubic N\'ed\'elec elements for $\boldsymbol{A}$ (provided by the built-in \textsc{AC/DC} module). 
We have typically employed in our dynamo models between $4 \times 10^5$ and $10^6$ degrees of freedom. 
PC BC (\ref{eq:BCsupra1}) are directly implemented in this form, whereas PV BC (\ref{eq:BCferro}) are converted into Neumann BC for the magnetic potential (which are built-in within \textsc{comsol}). 
We refer the reader to \cite{cebron2012magnetohydrodynamic} for further details about the finite-element implementation. 
To estimate the eigenvalue $\lambda = \sigma + \mathrm{i} \omega$, we solve the induction equation as an initial-value problem in time. 
The time-stepping uses the implicit differential-algebraic (IDA) solver, based on variable-coefficient backward differentiation formulae of variable order. 
We integrate the induction equation over at least one magnetic diffusion time, starting from an initial field $\boldsymbol{B}_0$.
We have checked that the dynamo computations are not affected by starting the simulations from other initial conditions for a few targeted runs. 
We then extract the growth rate $\sigma \geq 0$ of the fastest growing dynamo field (respectively the decay rate $\sigma < 0$ of the slowest decaying field), and the angular frequency $\omega$ of the corresponding mode, by post-processing the data as follows.
We adjust the time series of the volume-averaged magnetic energy $\langle \boldsymbol{B}^2 \rangle_V$ with the nonlinear fit
\begin{equation}
    \langle \boldsymbol{B}^2 \rangle_V = [a_0 + a_1 \sin^2 (\omega t + \Delta \phi)] \exp (2 \sigma t),
\end{equation}
where $[a_0,a_1]$ are adjustable amplitudes and $\Delta \phi$ is an adjustable phase lag.
This fitting procedure allows an accurate determination of $\sigma$ and $\omega$ when the integration time is long enough.

\begin{figure}
    \centering
    \begin{tabular}{cc}
        \includegraphics[width=0.47\textwidth]{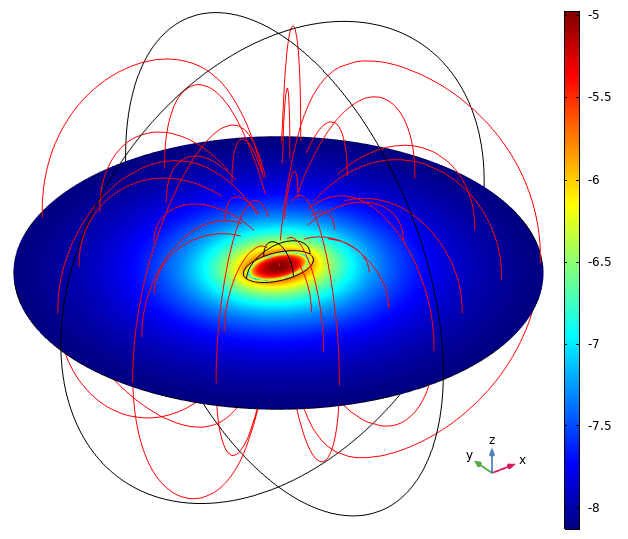} &
        \includegraphics[width=0.45\textwidth]{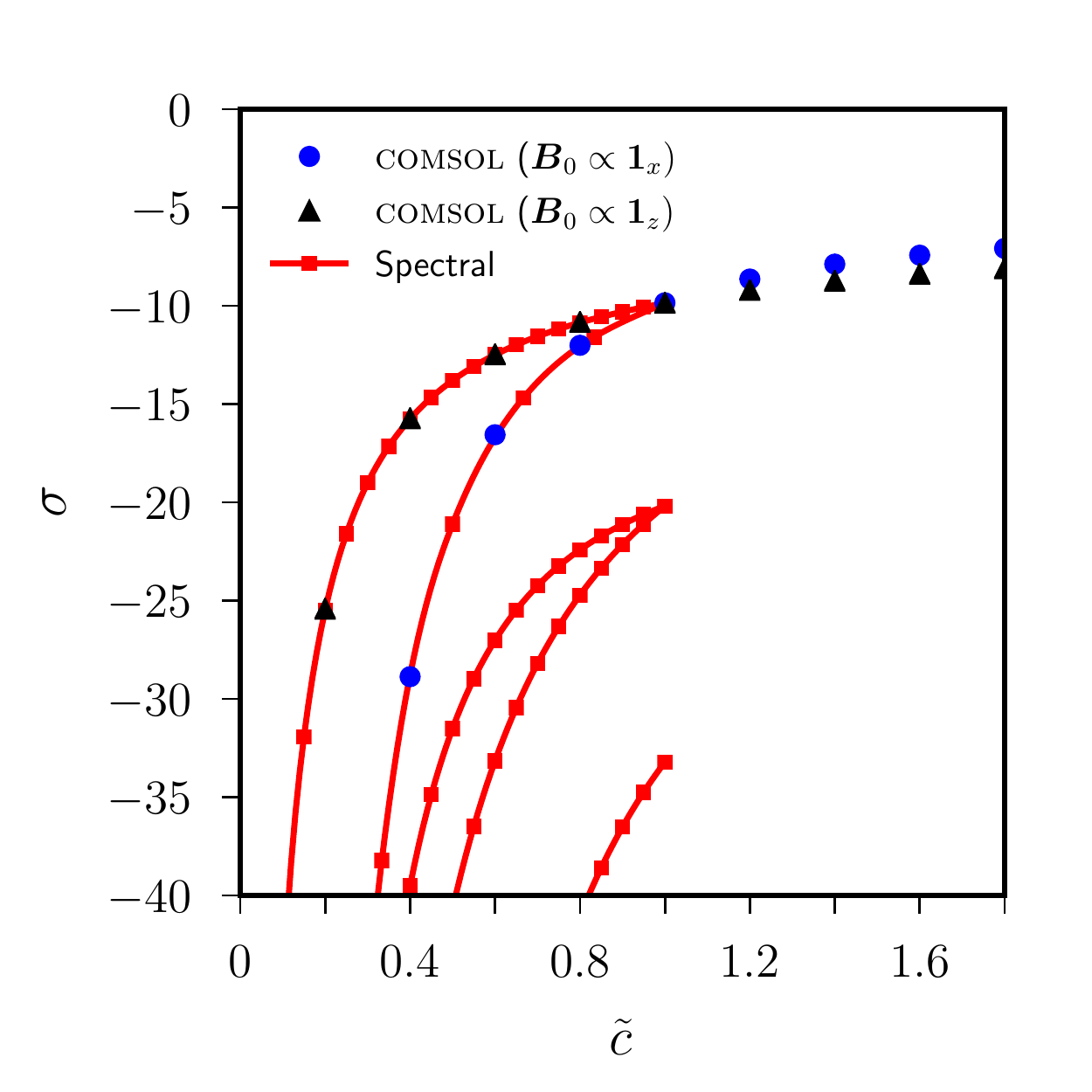} \\
        (a) & (b) \\
    \end{tabular}
    \caption{(a) Three-dimensional rendering of the slowest decaying magnetic mode for a triaxial ellipsoid with $\tilde{c}=0.8$ and $\beta=0.44$ (computed with \textsc{comsol} for an initial field $\boldsymbol{B}_0 \propto \boldsymbol{1}_z$). 
    Colour bar shows $\log_{10} ||\boldsymbol{B}||$, and a few magnetic field streamlines are shown in the exterior region. 
    The blue region illustrates the weakly conducting exterior region.
    (b) Dimensionless decay rate $\sigma$, as a function of dimensionless polar axis $\tilde{c}$, for the free-decay magnetic modes in spheroids ($\beta=0$).
    Comparison between \textsc{comsol} solutions for different initial fields $\boldsymbol{B}_0$, and non-polynomial spectral solutions with $\tilde{c} \leq 1$ (red squares, given by tables 1 and 2 in \cite{ivers2017kinematic}).
    Solid red curves show cubic interpolations from the red squares. (Online version in colour.)}
    \label{fig:ivers2017}
\end{figure}

The finite-element method can also be used to approximate IN BC as follows.
We assume that the fluid volume is surrounded by a motionless weakly conducting sphere (as illustrated in figure \ref{fig:ivers2017}a), with a large radius compared with the largest ellipsoidal semi-axis $a$.  
To model an insulating boundary, we integrate the induction equation (in dimensional form) $\partial_t \boldsymbol{A} = ( \mu \sigma_E)^{-1} \, \nabla^2 \boldsymbol{A}$ for the magnetic potential in the weakly diffusive (motionless) exterior, where $\sigma_E$ is the exterior electrical conductivity.
We assume $\sigma_E = 10^{-3} \sigma_f$ in our models, where $\sigma_f$ is the fluid electrical conductivity (ratios $\sigma_E/ \sigma_f = 10^{-4}-10^{-2}$ are usually adopted in DNS \cite{cebron2012magnetohydrodynamic,reddy2018turbulent}). 
When the distant boundary of the exterior region is far enough from the fluid domain, PV BC or PC BC can be enforced without changing the results in the fluid. 
We thus impose PC BC on a distant spherical boundary (here at the dimensional radius $r=8R$). 
To validate our approximate implementation of IN BC, we have computed in figure \ref{fig:ivers2017}b the free-decay magnetic modes in spheroids. 
We obtain an excellent quantitative agreement with the published solutions \cite{ivers2017kinematic}, although we have considered a weakly conducting exterior and not a perfect electrical insulator. 
As for the PV BC in figure \ref{fig:freedecay}b, different branches are numerically obtained for the slowest decaying modes (when starting from different initial conditions in spheroids).
Our finite-element approximation can thus be used for IN BC in ellipsoids.

\section{Polynomial elements for PV BC}
\label{appendix:basis}
\subsection{Admissible form}
We seek admissible polynomial elements of maximum degree $N$, which requires $\mathcal{A} \in \mathcal{P}_N$, $\mathcal{B} \in \mathcal{P}_{N-1}$, and $\Phi \in \mathcal{P}_{N+1}$ in decomposition (\ref{eq:spectralab}). 
Since $\mathcal{A}$ and $\Phi$ must vanish on the boundary, they are sought in the form $\mathcal{A}=(1-F) \widetilde{A}$ and $\Phi=(1-F) \widetilde{\Phi}$ with $F=(x/a)^2 + (y/b)^2 + (z/c)^2$, and where the two scalars $\widetilde{A} \in \mathcal{P}_{N-2}$ and $\widetilde{\Phi} \in \mathcal{P}_{N-1}$ do need to satisfy any BC. 
Then, if we parametrise the Cartesian coordinates $(x,y,z)$ by introducing the spherical-like coordinates $(\breve{r}, \theta, \phi)$ that map the ellipsoid into a computational sphere such that
\begin{subequations}
\begin{equation}
    x/a = \breve{x} = \breve{r} \sin \theta \cos \phi, \quad y/b = \breve{y} = \breve{r} \sin \theta \sin \phi, \quad z/c = \breve{z}= \breve{r} \cos \theta,
    \tag{\theequation a--c}
\end{equation}
\end{subequations}
the two scalars $[\widetilde{A},\widetilde{\Phi}]$ must admit the polynomial form $[\widetilde{A}, \widetilde{\Phi}] \propto \breve{r}^{2} \mathcal{H}_l^m \{ \breve{x}, \breve{y}, \breve{z} \}$ according to the spherical harmonic expansion theorem \cite{dassios2012ellipsoidal}, where $\breve{r}^2 = \breve{x}^2 + \breve{y}^2 + \breve{z}^2$ is the re-scaled radius, and $\mathcal{H}_l^m \{ \breve{x},\breve{y},\breve{z} \}$ are the solid spherical harmonics of degree $l\geq 1$ (the $l=0$ harmonic is excluded to have uniquely defined vector fields \cite{backus1996foundations}) and order $|m|\leq l$.
Powers in $\breve{r}^2$ guarantee enough differentiability of $\boldsymbol{B}$ at the centre \cite{dudley1989time,lewis1990physical}. 
As shown below, $\mathcal{H}_l^m \{ \breve{x},\breve{y},\breve{z} \}$ admits exact polynomial expansions in the re-scaled Cartesian coordinates $(\breve{x},\breve{y},\breve{z})$. 
To be compatible with the polynomial expansion in the spherical-like coordinates, the two scalars $[\mathcal{A}, \Phi]$ must therefore be expanded in the original Cartesian coordinates $(x,y,z)$ as
\begin{subequations}
\label{eq:ACscalarPOL}
\allowdisplaybreaks
\begin{align}
    \mathcal{A} &= (1-F) \sum_{l=1}^{N-2} \sum_{m=-l}^{l} \sum_{p=0}^{\lfloor \frac{1}{2} (N-l) \rfloor -1} \alpha_{p,l}^m \, F^p \, \mathcal{H}_l^m \left \{ \frac{x}{a}, \frac{y}{b}, \frac{z}{c} \right \}, \\
    \Phi  &= (1-F) \sum_{l=1}^{N-1} \sum_{m=-l}^{l} \sum_{p=0}^{\lfloor \frac{1}{2} (N+1-l) \rfloor -1} \phi_{p,l}^m \, F^p \, \mathcal{H}_l^m \left \{ \frac{x}{a}, \frac{y}{b}, \frac{z}{c} \right \}, 
\end{align}
\end{subequations}
where $[\alpha_{p,l}^m, \phi_{p,l}^m]$ are unknown coefficients. 
The scalar $\mathcal{B} \in \mathcal{P}_{N-1}$ is then obtained by solving equation (\ref{eq:diva}), which gives
\begin{equation}
    \mathcal{B} = \sum_{i+j+k\leq N-1} -\frac{\delta_{ijk}}{(i+1)/a^2 + (j+1)/b^2 + (k+1)/c^2} \,x^i y^j z^k
\end{equation}
where we have written $\nabla^2 \Phi \propto \delta_{ijk} \, x^i y^j z^k$ by using equation (\ref{eq:ACscalarPOL}b). 

\subsection{Cartesian form of solid spherical harmonics}
The solid spherical harmonics $\mathcal{H}_l^m \{\breve{x}, \breve{y}, \breve{z} \} = \breve{r}^l \mathcal{Y}_l^m$, where $\mathcal{Y}_l^m$ is the spherical harmonic of degree $l\geq 0$ and order $|m|\leq l$, admit explicit Cartesian expressions involving monomial of finite degree $\breve{x}^i \breve{y}^j \breve{z}^k$ \cite{backus1996foundations}.
We use here the real-valued Cartesian forms  (using Schmidt semi-normalisation)
\begin{equation}
    \begin{pmatrix}
    \mathcal{H}_l^m \{ \breve{x}, \breve{y}, \breve{z} \} \\
    \mathcal{H}_l^{-m} \{ \breve{x}, \breve{y}, \breve{z} \} \\
     \mathcal{H}_l^{0} \{ \breve{x}, \breve{y}, \breve{z} \} \\
    \end{pmatrix} = \sqrt{\frac{2l+1}{4 \pi} \frac{(l-m)!}{(l+m)!}} \begin{pmatrix}
    \Pi_l^m(\breve{r}^2, \breve{z}) \, A_m(\breve{x},\breve{y}) \\
    \Pi_l^m(\breve{r}^2, \breve{z}) \, B_m(\breve{x},\breve{y}) \\
    \Pi_l^0(\breve{r}^2, \breve{z}) \\
    \end{pmatrix}
    \label{eq:solidYlm}
\end{equation}
with
\begin{equation}
   [ A_m, \ B_m ] \, (\breve{x},\breve{y}) = \sum_{k=0}^{m} \begin{pmatrix} m \\ k \\ \end{pmatrix} \breve{x}^k \breve{y}^{m-k} \left [ \cos \left ((m-k)\frac{\pi}{2} \right ), \ \sin \left ((m-k)\frac{\pi}{2} \right ) \right ],
\end{equation}
and
\begin{subequations}
\begin{align}
    \Pi_l^m (\breve{r}^2,\breve{z}) &= \sum_{k=0}^{\lfloor (l-m)/2 \rfloor} (-1)^k 2^{-l} \begin{pmatrix} l \\ k \end{pmatrix} \begin{pmatrix} 2l - 2k \\ l \end{pmatrix} \frac{(l-2k)!}{(l-2k-m)!} \breve{r}^{2k} \breve{z}^{l-2k-m}, \\
    \Pi_l^0 (\breve{r}^2,\breve{z}) &= \sum_{k=0}^{\lfloor l/2 \rfloor} (-1)^k 2^{-l} \begin{pmatrix} l \\ k \end{pmatrix} \begin{pmatrix} 2l - 2k \\ l \end{pmatrix} \breve{r}^{2k} \breve{z}^{l-2k}.
\end{align}
\end{subequations}


{
\bibliography{./main}
\bibliographystyle{RS.bst}
}

\end{document}